\documentclass[twocolumn,showpacs,preprintnumbers,amsmath,amssymb,prb]{revtex4}

\usepackage{graphicx}
\usepackage{dcolumn}
\usepackage{bm}
\usepackage{amsmath,amssymb}

\newcommand{\be}{\begin{equation}}
\newcommand{\ee}{\end{equation}}

\newcommand{\beq}{\begin{eqnarray}}
\newcommand{\eeq}{\end{eqnarray}}

\def\eq#1{(\ref{#1})}
\def\H1{\widehat{H}_1}

\begin{document}

\title{Charge and spin density response functions of the clean two-dimensional 
electron gas with Rashba spin-orbit coupling at finite momenta and frequencies}

\author{M. Pletyukhov}

\author{S. Konschuh}
\affiliation{Institut f\"ur Theoretische Festk\"orperphysik and Center for 
Functional Nanostructures, Universit\"at Karlsruhe, D-76128 Karlsruhe, Germany}

\begin{abstract}
We analytically evaluate charge and spin density response functions of the 
clean two-dimensional electron gas with Rashba spin-orbit coupling at finite 
momenta and frequencies. On the basis of our exact expressions we discuss the 
accuracy of the long-wavelength and the quasiclassical approximations. We also 
derive the static limit of spin susceptibilities and demonstrate, 
in particular, how the Kohn-like anomalies in their derivatives are related 
to the spin-orbit modification of the Ruderman-Kittel-Kasuya-Yosida 
interaction. Taking into account screening and exchange effects of the Coulomb 
interaction, we describe the collective charge and spin density excitation 
modes which appear to be coupled due to nonvanishing spin-charge response 
function.
\end{abstract}
\maketitle

\section{Introduction}

One of the working principles of semiconductor spintronics\cite{wolf,ZFS} is 
based on the idea to exploit spin-orbit (SO) coupling for a manipulation of an
electron's spin by means of electric fields. The SO coupling of the Rashba 
type\cite{BR} arises in a two-dimensional electron gas (2DEG) at semiconductor 
heterojunction due to the quantum well asymmetry in the perpendicular 
direction, and the strength $\alpha_R$ of this coupling can be tuned by a gate 
voltage\cite{Nitta,Koga}.

A theoretical description of SO-related phenomena in the 2DEG is provided by 
coupled transport equations for  charge and spin components of the 
distribution function\cite{MH,MSH,burkov,malsh,schwab}. In the regime of a 
linear response to external fields these equations appear to be  intimately 
linked to  the density response functions such as charge and  spin 
susceptibilities and -- more peculiar -- spin-charge response functions. 
In the presence of impurity scattering all of these functions  have been 
previously evaluated in the quasiclassical approximation in the both 
diffusive\cite{burkov} ($q  \ll \frac{\hbar}{v_F \tau}$) and 
non-diffusive\cite{pletcr} ($\frac{\hbar}{v_F \tau} \ll q \ll k_F$) regimes, 
where $q$ is a momentum transfer, $k_F$ and $v_F$ are Fermi momentum and Fermi 
velocity, and  $\tau$ is an elastic scattering time. 

Recently it has been also remarked\cite{shnirman,pletcr} that the 
quasiclassical results for the Rashba system are validated only in the 
presence of a finite amount of disorder such that 
$\tau^{-1} \gg m^* \alpha_R^2 $, where $m^*$ is an effective electron's mass. 
Therefore, they  cannot be straightforwardly applied  in the extreme 
collisionless limit $\tau \to \infty$ even at small $q \ll k_F$, and the 
response functions of the clean 2DEG with Rashba SO coupling require a more 
refine consideration at finite values of $q$ and frequency $\omega$. 

Various response functions in the clean case $\tau \to \infty$ are most easily 
evaluated in the long-wavelength limit 
$q \to 0$\cite{MH,MCE,ras1,schliem,mont}. On the other hand, the knowledge of 
the dynamic response functions at finite $q$ enables one to find dispersions 
of the collective charge and spin density 
excitations\cite{brataas,ullrich,xu,wang,gumbs,kush1,pletgr} occurring in the 
presence of  electron-electron interaction. In particular, in 
Ref.~\onlinecite{pletgr} the polarization operator, or the charge 
susceptibility, of the system in question has been calculated analytically 
at arbitrary momenta and frequencies, and the SO-induced attenuation of the 
charge density mode (plasmon) has been quantitatively described within the 
random phase approximation (RPA). 

In this paper we analytically evaluate (Sec.~\ref{operator}) the other density 
response functions following the computational scheme elaborated in 
Ref.~\onlinecite{pletgr}. In Sec.~\ref{static} we derive the static limit of 
the spin susceptibilities, and observe an occurrence of the Kohn-like 
anomalies\cite{Stern} in their derivatives. We demonstrate how they are 
related to the SO modification\cite{bruno} of the  
Ruderman-Kittel-Kasuya-Yosida (RKKY) 
interaction\cite{rkkyref1,rkkyref2,rkkyref3} between two magnetic impurities. 
In Sec.~\ref{smallq} we compare our expressions for the dynamic response 
functions with the results of the long-wavelength and the quasiclassical 
approximations, and make conclusions about applicability ranges of the latter. 
Finally, in Sec.~\ref{collect} we revisit the problem of the collective charge 
and spin density excitations treating electron-electron interaction in terms 
of the Hubbard approximation\cite{Mahan}, i.e. our consideration extends 
beyond the RPA scheme. We demonstrate that the charge and the spin components 
are coupled in the obtained collective modes, which is a consequence of 
a simultaneous account of  the non-zero spin-charge response function and the 
exchange vertex corrections.

\section{Basic definitions}
\label{defin}
Let us consider the 2DEG with SO coupling of the Rashba type \cite{BR}
which is described by the single-particle Hamiltonian
\be
H_{{\bf k}} = \frac{k^2}{2m^{*}}+
\alpha_{R} k \, h^R_{{\bf k}},
\label{initham}
\ee
where  $h^R_{{\bf k}} = \sigma^x \sin \phi_{{\bf k}} -\sigma^y 
\cos \phi_{{\bf k}}$ is the spin-angular part of the Rashba spin-orbit 
coupling term, and we use  the units such that $\hbar=1$. 

The spectrum of \eq{initham} is split into two subbands,
\be
\epsilon_{{\bf k}}^{\pm} =\frac{k^2}{2 m^*} \pm  \alpha_{\mathrm{R}}k,
\label{SOspectrum}
\ee
the corresponding eigenstates being
\be
\psi_{{\bf k} +} = \frac{1}{\sqrt{2}} 
\left(\begin{array}{c} 1 \\ -i e^{i \phi_{{\bf k}}} \end{array} \right), 
\quad \psi_{{\bf k}-} = \frac{1}{\sqrt{2}} 
\left(\begin{array}{c} - i e^{-i \phi_{{\bf k}}} \\ 1 \end{array} \right).
\label{eigen}
\ee
The matrix ${\cal U}_{{\bf k}}$ diagonalizing the initial Hamiltonian 
\eq{initham} as well as $h^R_{{\bf k}}$, i.e. 
$\sigma^z = {\cal U}^{\dagger}_{{\bf k}} h^R_{{\bf k}} {\cal U}_{{\bf k}}$, 
is then given by 
\be
{\cal U}_{{\bf k}} = \left( \psi_{{\bf k} +} ; \psi_{{\bf k} -} \right) 
= \frac{1}{\sqrt{2}} [1 - i \sigma^x \cos \phi_{{\bf k}} 
- i \sigma^y \sin \phi_{{\bf k}}].
\label{ukm}
\ee
It is also convenient to introduce the projectors onto the eigenstates 
\eq{eigen},
\beq
P_{{\bf k} \pm} = \psi_{{\bf k}\pm} \otimes \psi_{{\bf k}\pm}^{\dagger} \equiv
\frac{1 \pm h^R_{{\bf k}}}{2},
\eeq
which allow us, for example, to spectrally decompose the Hamiltonian 
$H_{{\bf k}} = \sum_{\mu=\pm} \epsilon^{\mu}_{{\bf k}} P_{{\bf k} \mu}$ 
as well as the (retarded) Green's function
\be
G_{{\bf k} \epsilon}^{{\rm ret}} = \sum_{\mu=\pm} 
\frac{P_{{\bf k} \mu}}{\epsilon +i 0 -\epsilon_{{\bf k}}^{\mu}}.
\label{green}
\ee

A linear response $\rho^{\alpha}_{{\bf q} \omega}$ of the charge density 
($\alpha =0$) and the spin densities ($\alpha=x,y,z$) to an external spatially 
inhomogeneous and  nonstationary perturbation $V^{\beta}_{{\bf q} \omega}$, 
which consists of a scalar potential ($\beta=0$) and a magnetic field 
($\beta =x,y,z$), is usually determined in the framework of the Kubo 
formalism\cite{Mahan}. Applying the standard technique of the linear response 
theory and using for convenience the representation \eq{green}, one can 
establish an expression for the (retarded) density-density response functions,
\be
\chi_{{\bf q} \omega}^{\alpha \beta} = \sum_{\mu,\mu' =\pm}
\int \frac{d^2 {\bf k}}{(2 \pi)^2}\,\,
\frac{n_F (\epsilon_{{\bf k}}^{\mu}) 
-n_F (\epsilon_{{\bf k}+{\bf q}}^{\mu'})}
{\omega+ i 0
+ \epsilon_{{\bf k}}^{\mu} - \epsilon_{{\bf k}+{\bf q}}^{\mu'}}
{\cal F}^{\alpha \beta}_{{\bf k}, {\bf k}+{\bf q}; \mu, \mu' },
\label{chidef}
\ee
where
\be
{{\cal F}}^{\alpha \beta}_{{\bf k}, {\bf k}+{\bf q}; \mu, \mu'} =
{\rm Tr} [ P_{{\bf k}\mu} \sigma^{\alpha} P_{{\bf k}+{\bf q},\mu'} 
\sigma^{\beta}]
\label{overlap}
\ee
are the overlap functions;  $n_F$ denotes the Fermi distribution, and 
$\sigma^0 \equiv 1$. 

In the explicit form  ${{\cal F}}^{\alpha \beta}_{{\bf k}, {\bf k}+{\bf q}; 
\mu, \mu'}$ are listed in Appendix B of Ref.~\onlinecite{pletcr}. Here we 
quote their symmetry property, which can be directly established from the 
definition \eq{overlap}:
\be
{\cal F}^{\alpha \beta}_{{\bf k},{\bf k}+{\bf q};\mu, \mu'} 
= s {\cal F}^{\alpha \beta}_{-{\bf k}-{\bf q},-{\bf k}; \mu', \mu},
\label{symref}
\ee 
where $s=1$ for the charge-charge and spin-spin components, and $s=-1$ for the
spin-charge components.

The expression \eq{chidef} includes definitions of a polarization operator 
($\alpha, \beta =0$), spin susceptibilities ($\alpha, \beta =x,y,z$) as well 
as of spin-charge response functions ($\alpha =0$ and $\beta =x,y,z$, or vice 
versa). In the presence of SO coupling the latter functions do not vanish, and 
their study represents  an especial interest. 

We note that the expression \eq{chidef} can be alternatively found in terms of 
the equations of motion for the local charge and spin densities (see Appendix 
\ref{eom} for details). A matrix formulation of this approach provides a 
convenient tool for an account of screening and exchange effects in presence 
of electron-electron interaction. In more detail this will be discussed in 
Sec.~\ref{collect}.

\section{Evaluation of $\chi^{\alpha \beta}$}
\label{operator}

The functions \eq{chidef} have been previously treated at finite $q$ in terms 
of different approximations. The most typical of them are: 1) the small-$q$ 
(long-wavelength) formal expansion of the whole integrand (see, e.g., 
Refs.~\onlinecite{MH,MCE}); and 2) the quasiclassical approximation (see, e.g.,
 Ref.~\onlinecite{brataas}) which is usually performed in the quantum kinetic 
equation approach.  

In Ref.~\onlinecite{pletgr} the polarization operator $\chi^{00}$ in the clean 
limit has been evaluated beyond these approximations, and the obtained result 
has been used for an estimation of  the accuracy of the long-wavelength 
expansion. In particular, the latter has been shown to be applicable in the 
limited range of the very small $q \ll k_R^2/k_F$, where 
$k_F= \sqrt{2 m^* \epsilon_F + k_R^2}$ and $k_R =m^* \alpha_R$ is the Rashba 
momentum splitting.

As for the quasiclassical approximation, it has been argued in 
Ref.~\onlinecite{pletcr} that its application in the presence of SO coupling 
is validated at the  finite values of the disorder broadening 
$\tau^{-1} \gg k_R^2/m^*$, which smoothens the divergences of the 
quasiclassical result near the boundaries of the SO-induced (intersubband) 
particle-hole excitation region in the $(q,\omega)$-plane. One still might 
hope that the quasiclassical approximation is trustful  in the extreme 
collisionless limit $\tau \to \infty$, provided one does not come too close 
to the boundaries in question. For this reason we are going to revisit its 
accuracy in the context of our present calculations.
 
It also remains unclear how the two above mentioned approximations are related 
to each other in the clean limit. Both of them are elaborated for small values 
of $q$, but seem to give different results even at $q \to 0$. For example, an 
application  of the long-wavelength expansion to the  optical conductivity 
yields a box-like function\cite{MCE,ras1} which is finite at frequencies 
$2 \alpha_R k_F - 2 m^* \alpha_R^2 < \omega <  2 \alpha_R k_F + 2 m^* 
\alpha_R^2$, while in the quasiclassical approximation the width of this 
frequency window cannot be resolved at all.

In this section we are going to evaluate the functions \eq{chidef} without 
making any kind of approximations. We will neatly follow the computational 
scheme elaborated in Ref.~\onlinecite{pletgr} for $\chi^{00}$. Thus, we will 
derive analytic expressions for the other response functions and discuss  
their limiting behavior in the subsequent sections.

We start from the observation that, due to the  momentum space isotropy of the 
spectrum \eq{SOspectrum}, the response functions \eq{chidef} can be 
represented in the form
\beq
& & \chi_{{\bf q} \omega}^{\alpha \beta} = \sum_{\mu,\lambda =\pm} 
\chi_{{\bf q} \omega, \lambda}^{\alpha \beta, \mu} , \\
& & \chi_{{\bf q} \omega, \lambda}^{\alpha \beta, \mu}=  
\int \frac{d^2 {\bf k}}{(2 \pi)^2} n_F (\epsilon_{{\bf k}}^{\mu}) 
\lambda^{\frac{1-s}{2}} \\
& & \times \left[ \frac{{\cal F}^{\alpha \beta}_{{\bf k},{\bf k}+{\bf
q}; \mu, \mu}}{\epsilon_{{\bf k}}^{\mu} - \epsilon_{{\bf k}+{\bf q}}^{\mu} +
\lambda (\omega + i 0)} +  \frac{{\cal F}^{\alpha \beta}_{ {\bf k},{\bf k}
+{\bf q}; \mu, - \mu}}{\epsilon_{{\bf k}}^{\mu} -  \epsilon_{{\bf k}+{\bf
q}}^{-\mu} + \lambda (\omega + i 0)} \right], \nonumber
\eeq
where the factor $\lambda^{\frac{1-s}{2}}$ originates from the property 
\eq{symref}. It is convenient to choose the basis in the momentum space 
such that ${\bf q}$ is aligned with $x$-direction. Then the matrix $\chi$ 
becomes sparse, the non-vanishing terms being $\chi^{00}$, $\chi^{xx}$, 
$\chi^{yy}$, $\chi^{zz}$, $\chi^{0y} = \chi^{y0}$, and $\chi^{zx} 
= -\chi^{xz}$. Introducing $x = \cos (\phi_{{\bf k}} - \phi_{{\bf q}}) 
\equiv \cos \phi$, we obtain the following expressions at zero temperature
\beq
& & \left( \begin{array}{c} \chi_{{\bf q} \omega, \lambda}^{00,\mu} \\  
\chi_{{\bf q} \omega, \lambda}^{zz,\mu} \end{array} \right) = 
\frac{1}{8 \pi^2} 
\int_0^{k_F - \mu k_R} k d k \int_0^{2 \pi} d \phi
\label{pilm00}\\
& & \times \left[ \frac{1 \pm \frac{k+q x}{|{\bf k} + {\bf q}|}}
{\epsilon_{{\bf k}}^{\mu} - \epsilon_{{\bf k}+ {\bf q}}^{\mu} +
\lambda (\omega+ i 0)} 
+ \frac{1\mp \frac{k+q x}{|{\bf k} + {\bf q}|}}
{\epsilon_{{\bf k}}^{\mu} - \epsilon_{{\bf k}+ {\bf q}}^{-\mu} + 
\lambda (\omega + i 0)} \right]  ,
\nonumber
\eeq
\beq
& & \left( \begin{array}{c} \chi_{{\bf q} \omega, \lambda}^{yy,\mu} \\  
\chi_{{\bf q} \omega, \lambda}^{xx,\mu} \end{array} \right) =
\frac{1}{8 \pi^2} 
\int_0^{k_F - \mu k_R} k d k \int_0^{2 \pi} d \phi
\label{pilmyy}\\
& & \times \left[ \frac{1 \pm \frac{k (2 x^2 -1)+q x}{|{\bf k} + {\bf q}|}}
{\epsilon_{{\bf k}}^{\mu} - \epsilon_{{\bf k}+ {\bf q}}^{\mu} +
\lambda (\omega+ i 0)} 
+ \frac{1\mp \frac{k (2 x^2 -1) +q x}{|{\bf k} + {\bf q}|}}
{\epsilon_{{\bf k}}^{\mu} - \epsilon_{{\bf k}+ {\bf q}}^{-\mu} + 
\lambda (\omega + i 0)} \right]  ,
\nonumber
\eeq
\beq
& & \left( \begin{array}{c} \chi_{{\bf q} \omega, \lambda}^{0y,\mu} \\  
i \chi_{{\bf q} \omega, \lambda}^{zx,\mu} \end{array} \right) =
\frac{\mu (-\lambda)^{\frac{1 \pm 1}{2}}}{8 \pi^2} 
\int_0^{k_F - \mu k_R} k d k \int_0^{2 \pi} d \phi \,\,\, 
\label{pilm0y}\\
& & \times \left[ \frac{x \pm \frac{q+k x}{|{\bf k} + {\bf q}|}}
{\epsilon_{{\bf k}}^{\mu} - \epsilon_{{\bf k}+ {\bf q}}^{\mu} +
\lambda (\omega+ i 0)} 
+ \frac{x\mp \frac{q+k x}{|{\bf k} + {\bf q}|}}
{\epsilon_{{\bf k}}^{\mu} - \epsilon_{{\bf k}+ {\bf q}}^{-\mu} + 
\lambda (\omega + i 0)} \right]  .
\nonumber
\eeq
Note that the components of $\chi$ in the arbitrary basis can be recovered 
by an orthogonal rotation in the $x-y$ plane (see Appendix B of 
Ref.~\onlinecite{pletcr}).

After simple algebra we eliminate the  odd powers of $|{\bf k} +{\bf q}|$ 
in Eqs. \eq{pilm00}-\eq{pilm0y}. It means that the corresponding integrands 
happen to be  rational functions of $k$ and $\cos \phi$. Let us also note the 
identity
\be
\chi^{xx} +\chi^{yy} = \chi^{00}+\chi^{zz},
\label{iden}
\ee
which allows us to express, say, $\chi^{xx}$ in terms of the other diagonal 
components. There remain, in fact, only  five independent functions 
$\chi^{(j)} = \{ \chi^{00},\chi^{0y},\chi^{yy},\chi^{zz} , i \chi^{zx} \}$, 
which can be conveniently labeled by the index $j=1, \ldots,5$.  Like in 
Ref.~\onlinecite{pletgr}, we also introduce the index $i=1,2,3,4$ which 
denotes different combinations of  $\{\mu, \lambda \} 
= \{ -,+\},\{ +,+\},\{ -,-\},\{ +,-\}$, respectively. 

Defining the dimensionless units $y= k_R / k_F$, $z=q / 2 k_F$, $v= k/ k_F$, 
and $w=m^* \omega/2 k_F^2$, we cast \eq{pilm00}-\eq{pilm0y} into the form
\beq
- \frac{1}{\nu} {\rm Im}\chi_i^{(j)} &=& \int_0^{1
-\mu y} v g_i^{(j)} (v,z, w, y) d v ,\label{pregunf} \\
- \frac{1}{\nu} {\rm Re} \chi_i^{(j)} &=& \check{f}_i^{(j)} + \int_0^{1
-\mu y} v f_i^{(j)} (v,z, w, y) d v ,
\label{prefunf}
\eeq
where $\nu \equiv \nu_{2D} = \frac{m^*}{2 \pi}$ is the density of states
in 2DEG per each spin component. The functions $g_i^{(j)}$ and $f_i^{(j)}$ are 
given by
\beq
g_i^{(j)} &=& \frac{\lambda C_i^{(j)}}{2} \int_{0}^{2 \pi} d \phi  \,\, 
{\rm sign} (2 v z x - \mu y v + 2 (z^2 -\lambda w)) \nonumber \\
& & \qquad \times  (x+ \delta_i^{(j)}) \,\, \delta (x^2 +\beta_i x +\gamma_i),
\label{gunf} \\
f_i^{(j)} &=& \frac{C_i^{(j)}}{2 \pi} \int_0^{2 \pi} d \phi
\frac{x + \delta_i^{(j)}}{x^2 + \beta_i x + \gamma_i},
\label{funf}
\eeq
where the coefficients 
\beq
\beta_i &=& \frac{2 (z^2-\lambda w) - \mu y (v +\mu y)}{v z}, \\
\gamma_i &=& 
\frac{(z^2 - \lambda w)^2 -\mu y v (z^2 - \lambda w) -z^2 y^2}{v^2 z^2},
\label{ingred}
\eeq
are the same for each $j$, and  $\delta (\ldots)$ in \eq{gunf} denotes the 
Dirac delta function. The difference between the response functions 
$\chi^{(j)}$ appears only in  the form of the coefficients 
$\check{f}_i^{(j)}$, $C^{(j)}_i$, and $\delta^{(j)}_i$, which are listed in 
Appendix \ref{toucal} for all $j$'s.

We note that the real part of $\chi^{(j)}$ can be represented as a sum 
\be
{\rm Re} \chi^{(j)} = \check{\chi}^{(j)} + {\rm Re} \chi^{(j),I}+ 
{\rm Re} \chi^{(j),II},
\ee
where the term $\check{\chi}^{(j)}= -\nu \sum_i \check{f}_i^{(j)}$ is nonzero 
only for $j=3$ and $j=5$ [see Eq. \eq{cheq}]. The terms 
${\rm Re} \chi^{(j), I}$ and ${\rm Re} \chi^{(j),II}$ are obtained by 
integrating the functions $\sum_i v f_i^{(j),I}$ and $\sum_i v f_i^{(j),II}$, 
where $f_i^{(j),I} = f_i^{(j)} \Theta (\beta_i^2 -4 \gamma_i)$ and 
$f_i^{(j),II} = f_i^{(j)} \Theta (4 \gamma_i -\beta_i^2)$.

Performing angular integration and a subsequent change of the  variable 
$v \to \tau (v)$ according to  Eq. (34) of  Ref.~\onlinecite{pletgr}, we 
obtain expressions for ${\rm Im} \chi^{(j)}$ and ${\rm Re} \chi^{(j),I}$ 
in the form of $\chi^{(1)}$ found previously,
\beq
&-& \frac{1}{\nu} {\rm Im} \chi^{(j)}  = \sum_{\sigma,\mu= \pm} \sigma 
\int_{\tau_{\sigma +} (y)}^{\tau_{\sigma +} (\mu)} 
d \tau {\cal L}^{(j)+} (\tau) 
\nonumber  \\
& & + \Theta (1 - 4 w) \sum_{\mu=\pm} \int_{\tau_{+-} (\mu)}^{\tau_{--} (\mu)} 
d \tau {\cal L}^{(j)-} (\tau) \label{impi} \\
& & + 2 \Theta (y^2 - 4 w) \int_{\tau_{--}(y)}^{\tau_{+-} (y)} 
d \tau {\cal L}^{(j)-} (\tau) , \nonumber
\eeq
and
\beq
&-& \frac{1}{\nu} {\rm Re} \chi^{(j),I} =  \sum_{\sigma,\mu=\pm} 
\int_{\tau_{\sigma +} (y)}^{\tau_{\sigma +} (\mu)} d \tau {\cal R}^{(j)+} 
(\tau) \nonumber \\
& & + \Theta (1 - 4 w) \sum_{\sigma, \mu=\pm} 
\int_{- \mu \tau_{++} (0)}^{\tau_{\sigma -} (\mu)} d \tau {\cal R}^{(j)-} 
(\tau) \label{realpi} \\
& & + 2 \Theta (y^2 - 4 w) \sum_{\sigma=\pm} 
\int_{\tau_{\sigma -}(y)}^{\tau_{-+} (0)} d \tau {\cal R}^{(j)-} (\tau) , 
\nonumber
\eeq
where
\beq
\tau_{1,2} &=& \pm w/z, \quad \tau_{3,4} = -y \pm z ,
\label{tauk} \\
\tau_{\sigma \lambda} (x) &=& 
\frac12 \left[-x +\sigma \sqrt{x^2 + 4 \lambda w} \right], \\
{\cal L}^{(j) \pm} (\tau) &=& {\cal L}^{(j)} (\tau) \,\, 
{\rm sign} (\tau^2 + y \tau \pm w), \\
{\cal R}^{(j) \pm} (\tau) &=& {\cal R}^{(j)} (\tau) \,\, 
{\rm sign} (\tau z \mp w (\tau +y)/z). 
\eeq

In the representation \eq{impi}-\eq{realpi} the actual integration limits are 
universal for all response function. The difference appears only in the form 
of the integrands
\beq
{\cal L}^{(j)} (\tau)&=& Q^{(j)} (\tau)  \frac{\Theta 
\left(P (\tau )\right)}{\sqrt{P (\tau)}}, \label{lt} \\
{\cal R}^{(j)} (\tau) &=& Q^{(j)} (\tau) \frac{\Theta 
\left(- P (\tau )\right)}{\sqrt{-P (\tau)}}, \label{rt} \\
P (\tau) &=& \prod_{k=1}^4 (\tau - \tau_k), 
\label{pt}
\eeq
which are specified for each response function by 
\beq
Q^{(1)} (\tau) &=& \frac{1}{2 z} (\tau -\tau_3) (\tau -\tau_4), \label{q1} \\
Q^{(2)} (\tau) &=& -\frac{w}{z \tau} Q^{(1)} (\tau), \label{q2} \\
Q^{(3)} (\tau) &=& \frac{w^2}{z^2 \tau^2} Q^{(1)} (\tau), \label{q3} \\
Q^{(4)} (\tau) &=& \frac{z}{2 \tau^2} (\tau -\tau_1 ) (\tau -\tau_2), 
\label{q4} \\
Q^{(5)} (\tau) &=& \frac{\tau+y}{z} Q^{(4)} (\tau). \label{q5}
\eeq
Interestingly, for the function $\chi^{xx} = \chi^{00}-\chi^{yy}+\chi^{zz}$ 
one would obtain the term
\be
Q^{(1)} (\tau) - Q^{(3)} (\tau) + Q^{(4)} (\tau) = \frac{(\tau+y)^2}{z^2} 
Q^{(4)} (\tau),
\ee
which is anticipated after comparison of Eqs. \eq{q4} and \eq{q5} with 
\eq{q1}-\eq{q3}.

Let us make several comments about the obtained results \eq{impi} and 
\eq{realpi}.

1) First of all, note that the overall sign in the second line of \eq{impi} 
differs from its counterpart in the corresponding equation (35) of 
Ref.~\onlinecite{pletgr}. We use the present opportunity to correct the 
misprint in the previously derived expression. Fortunately, it did not affect 
any other result of   Ref.~\onlinecite{pletgr}.

2) The actual intervals of integration in Eq.~\eq{impi} are explicitly 
written down in Eqs.~\eq{icha}-\eq{ichd}. None of them contains the 
point $\tau=0$, which means that one should not  worry about the convergence
of integrals over ${\cal L}^{(3,4,5)} (\tau) \sim 1/\tau^2$ near this point.

3) The explicit analytic relations for ${\rm Im} \chi^{(j)}$ in terms of 
elliptic functions\cite{gradst} can be found for all $j$'s in the same fashion 
as it has been done before for $\chi^{(1)}$ [see Appendix C of 
Ref.~\onlinecite{pletgr}].

4) Some of the actual integration intervals in  Eq.~\eq{realpi} do contain 
the point $\tau =0$, which means that the corresponding integrals  
$\int d \tau {\cal R}^{(3,4,5)} (\tau )$ are divergent in its vicinity. 
However, the whole expression Eq.~\eq{realpi} is convergent and well-defined, 
since the singularities exactly cancel each other. In order to make 
Eq.~\eq{realpi} practically useful, one has to substitute  
$\int d \tau {\cal R}^{(j)} (\tau )$ by the corresponding difference of 
primitives, which can be also found in terms of elliptic functions.

5)  Eq.~\eq{realpi} contains only the contribution ${\rm Re} \chi^{(j),I}$ 
to the full function ${\rm Re} \chi^{(j)}$. The contribution 
$\check{\chi}^{(j)}$ is quoted in  \eq{cheq}, and it remains to calculate 
the contribution ${\rm Re} \chi^{(j),II}$. Making a complex change of 
variables $\tau = \frac12 [-\mu (v+\mu y)+ i \sqrt{4 w - (v+\mu y)^2}]$ 
(cf. Eq. (34) of Ref.~\onlinecite{pletgr}) one can as well find  
${\rm Re} \chi^{(j),II}$ in the analytic form which would involve the 
same integrand $ {\cal R}^{(j)} (\tau )$ and a path of integration lying 
in the complex $\tau$-plane. Omitting technical details of this evaluation, 
we present  the  explicit expressions for ${\rm Re} \chi^{(j),I}+ 
{\rm Re} \chi^{(j),II}$ in Eqs.~\eq{rcha}-\eq{rchd}.

\section{Static limit}
\label{static}

In order to find a static limit of $\chi^{\alpha \beta}$ one should consider 
with caution Eqs. \eq{rcha} and \eq{rchd} at $w \to 0$. One can then find
\beq
&-& \nu^{-1} \lim_{w \to 0} \chi^{00} = 2 + \frac{\pi}{2} \Theta (y -|z-1|) 
\sin \psi \nonumber \\
&-& \sum_{\mu=\pm} \Theta (z-(1-\mu y)) \left(\mu \psi_{\mu} \sin \psi + 
\cos \psi_{\mu} + 2 l_{\mu} \cos \psi \right) \nonumber \\
& &   - 2 \Theta (z-1) \cos \psi \, \, {\rm arccosh} z , \label{s00} \\
&-&  \nu^{-1} \lim_{w \to 0} \chi^{zz} =2  + \frac{2}{\cos \psi} 
\sum_{\mu=\pm} \Theta (z-(1-\mu y)) l_{\mu}\nonumber \\
& &  + \frac{2 \Theta (z-1)}{\cos \psi}\left({\rm arccosh} z - 
\sqrt{1-(1/z)^2} \right), \label{szz} \\
&-&  \nu^{-1} \lim_{w \to 0} \chi^{yy}=  2 - 2 \Theta (z-1) \cos \psi 
\sqrt{1-(1/z)^2}, \label{syy} \\
&-&  i \nu^{-1} \lim_{w \to 0} \chi^{zx} =  \frac{\pi}{2} 
\Theta (y -|z-1|) 
\nonumber \\
& & - \sum_{\mu=\pm} \Theta (z-(1-\mu y)) \left(\mu \psi_{\mu} 
- 2 l_{\mu} \tan \psi  \right)  \nonumber \\
& & + 2 \Theta (z-1) \tan \psi \left({\rm arccosh} z -\sqrt{1-(1/z)^2} \right),
\label{szx} 
\eeq
where $\sin \psi = y/z$ (for $y<z$), $\sin \psi_{\mu} = (1 -\mu y)/z$ 
(for $1 -\mu y <z$; note that this notation differs from its counterpart 
in Ref.~\onlinecite{pletgr} by $\mu \to -\mu$), and 
\be
l_{\mu} = l_{\mu} (z) = \ln \frac{1 + z\sin (\psi_{\mu} + \mu \psi)}{2 
\sqrt{2 z} \cos \frac12 \psi_{\mu} \cos \frac12 \psi}.
\ee
The static limit of $\chi^{xx}$ can be found from the identity \eq{iden}. 
The off-diagonal spin-charge term $\chi^{0y}$ identically vanishes in this 
limit, which means a decoupling of charge and spin components at zero 
frequency.

\begin{figure}[t]
\includegraphics[width =8.25cm]{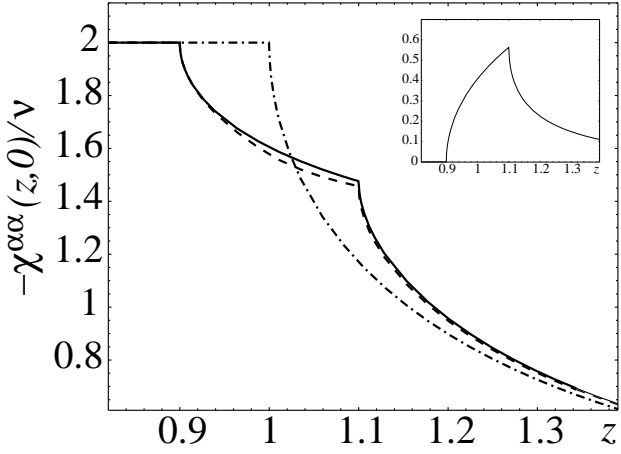}
\caption{Diagonal components of the static spin susceptibility near 
$z=q/2 k_F \sim 1$ plotted at $y=k_R /k_F =0.1$. Solid, dashed-dotted, 
and dashed lines correspond to $\alpha =x$, $\alpha =y$, and $\alpha =z$, 
respectively. The inset shows the off-diagonal component 
$- i \chi^{zx} (z,0)/\nu$. }
\label{statm}
\end{figure}

On the basis of the derived expressions \eq{s00}-\eq{szx} one can observe 
that for $z \leq 1-y$ all diagonal terms are equal to $ \chi^{\alpha \alpha}  
=- 2 \nu $, while their large-$z$ asymptotes are $ \chi^{\alpha \alpha} 
\approx - \nu \frac{1+y^2}{z^2}$. The off-diagonal spin-spin term $\chi^{zx}$ 
equals zero at $z \leq 1-y$, and $\chi^{zx} \approx \frac{2 i \nu y}{z^3} 
\left(1 + \frac{2 y^2}{3} \right)$ at large $z \gg 1$. The behavior of all 
components of the static spin susceptibility near $z \sim 1$ is shown in 
Fig.~\ref{statm}. One can see that $\chi^{yy}$ has a discontinuous derivative 
at $z=1$, while  the derivatives of $\chi^{xx}$, $\chi^{zz}$, and $\chi^{zx}$ 
are discontinuous at $z=1\mp y$. These anomalies are analogous to the Kohn 
anomaly of the polarization operator\cite{Mahan,Stern,pletgr} at $z=1$.

Using \eq{szz}-\eq{szx} we can find a SO-modification of the 
Ruderman-Kittel-Kasuya-Yosida (RKKY) 
Hamiltonian\cite{rkkyref1,rkkyref2,rkkyref3}, which describes an indirect 
exchange interaction  between two localized magnetic impurities. In general 
case the RKKY Hamiltonian reads\cite{bruno}
\be
H^{RKKY}_{1,2} = J^2_{RKKY} \sum_{\alpha,\beta=x,y,z} S_1^{\alpha} 
\chi^{\alpha \beta} ({\bf r}_{12})   S_2^{\beta},
\label{rkky0}
\ee
where ${\bf S}_{1,2}$ are the spin operators of impurities, and 
${\bf r}_{12} = {\bf r}_1 -{\bf r}_2$ is the distance between them. 

In the presence of SO coupling the Hamiltonian \eq{rkky0} becomes anisotropic 
in spin space, since the matrix $\chi$ is no longer proportional to the unit 
matrix. Let us find asymptotic values of $\chi^{\alpha \beta} ({\bf r}_{12})$  
at large $r_{12} \gg k_F^{-1}$. For simplicity we assume that the vector 
${\bf r}_{12}$ is aligned with $x$-direction in the coordinate space. 
Inspecting \eq{szz}-\eq{szx} and restoring the dimensional units, we establish 
the asymptotic form of the right-sided derivatives
\beq
\frac{d \chi^{yy}}{d q} \bigg|_{q \to q_{c 0}^+} & \approx & \frac{\nu}{2 k_F} 
\sqrt{\frac{2 q_{c0}}{q - q_{c0}}}, \\
\frac{d  \chi^{zz}}{d q}  \bigg|_{q \to q_{c \mu}^+} & \approx & 
\frac{d \chi ^{xx}}{d q}  \bigg|_{q \to q_{c \mu}^+} \approx \frac{\nu}{4 k_F} 
\sqrt{\frac{2 q_{c\mu}}{q - q_{c\mu}}}, \\
\frac{d \chi^{zx}}{d q} \bigg|_{q \to q_{c \mu}^+} & \approx & 
\frac{i \mu \nu}{4 k_F} \sqrt{\frac{2 q_{c\mu}}{q - q_{c\mu}}}
\eeq
near the discontinuity points $q_{c0}= 2 k_F$ and $q_{c\mu}= 2 k_F -2 \mu k_R$.
Performing the Fourier transformation of $\chi^{\alpha \beta} (q)$, we obtain 
the following leading asymptotic terms 
\beq
\chi^{yy} (r_{12}) & \approx & -\frac{\nu}{\pi} 
\frac{\sin (2 k_F r_{12})}{r_{12}^2}, \\
\chi^{zz} (r_{12} ) & \approx & \chi^{xx} (r_{12} ) \approx 
- \frac{\nu}{2 \pi} \sum_{\mu = \pm} \frac{q_{c \mu} 
\sin (q_{c \mu} r_{12})}{2 k_F r_{12}^2} \nonumber \\
& \approx & - \frac{\nu}{\pi} \frac{\sin (2 k_F r_{12})}{r_{12}^2} 
\cos (2 k_R r_{12} ), \\
\chi^{zx} (r_{12} ) & \approx &  - \frac{\nu}{2 \pi} \sum_{\mu = \pm} 
\mu \frac{q_{c \mu} \cos (q_{c \mu} r_{12})}{2 k_F r_{12}^2} \nonumber \\
& \approx & - \frac{\nu}{\pi} \frac{\sin (2 k_F r_{12})}{r_{12}^2} 
\sin (2 k_R r_{12} ),
\eeq
which oscillate in the coordinate space with the periods 
$2 \pi/ q_{c0}$ and $2 \pi / q_{c \mu}$. Substituting them into 
Eq.~\eq{rkky0}, we obtain
\be
H^{RKKY}_{1,2} = F_2 (r_{12}) \sum_{\alpha,\beta=x,y,z} S_1^{\alpha} 
O^{\alpha \beta} (\theta_{12})   S_2^{\beta},
\label{rkky1}
\ee 
where the range function
\be
F_2 (r_{12}) = -  J^2_{RKKY} \frac{\nu}{\pi} 
\frac{\sin (2 k_F r_{12})}{r_{12}^2}
\ee 
is the same as in the absence of SO coupling. The SO-modification of the 
Hamiltonian \eq{rkky0} consists in the spin twist determined by the orthogonal 
transformation
\be
O (\theta_{12} )= \left( \begin{array}{cccc} 
\cos \theta_{12}  & 0 & - \sin \theta_{12}  \\
0 & 1 & 0\\
\sin \theta_{12} & 0 & \cos \theta_{12}  
\end{array}\right)
\ee
with the rotation angle $\theta_{12} = 2 k_R r_{12}$. 
The expression \eq{rkky1} is in agreement with the corresponding result 
of Ref.~\onlinecite{bruno}.

\section{Behavior of the response functions $\chi^{\alpha \beta}$
at small momenta}
\label{smallq}

\subsection{Exact expressions for ${\rm Im} \chi^{\alpha \beta}$ in the 
SO-induced particle-hole excitation region}
\label{exactq}

\begin{figure}[b]
\includegraphics[width =8.25cm]{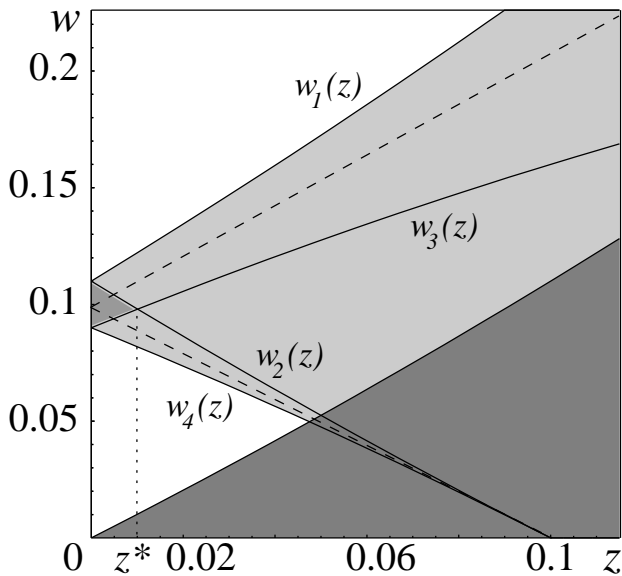}
\caption{SO-induced extension of the particle-hole excitation region at 
$y=0.1$ -- the light-gray area bounded by the parabolas $w_1 \equiv w_+$ 
and $w_4 \equiv w_-$. The small darkened triangle inside it indicates the 
range of applicability of the long-wavelength approximation. The dashed 
lines depict quasiclassical boundaries $w_{\pm}^{qc} = y \pm z$. }
\label{zstar}
\end{figure}

It has been discussed in Ref.~\onlinecite{pletgr} that an account of the 
Rashba  SO coupling leads to an extension of the boundaries of a particle-hole 
continuum, or Landau damping region, which is defined by ${\rm Im} \chi^{(1)} 
\neq 0$.  It has been also established that this extension has a shape of the 
wedge bounded by the parabolas 
$-(z-y)^2-(z-y) \equiv w_4 (z) < w < w_1 (z) \equiv (z+y)^2 + (z+y)$ 
[see Fig.~\ref{zstar}].

Since the representation \eq{pilm00}-\eq{pilm0y} manifests the same pole 
structure for all response functions, their imaginary parts appear to be 
nonzero in the same domain where ${\rm Im} \chi^{(1)} \neq 0$. Analyzing 
\eq{ichc}, we extract an explicit expression for ${\rm Im} \chi^{(j)}$ in 
the SO-induced particle-hole excitation region
\beq
&-& \frac{1}{\nu} {\rm Im} \chi^{(j)} = - \Theta (w_2 -w) 
\Theta (w - w_4) A_{2c}^{(j)} (z-y)\nonumber \\
& & -\Theta (w_1 -w) \Theta (w - w_2 ) A_{2c}^{(j)} (-t_1)\nonumber \\
& & +\Theta (w_3 -w) \Theta (w - w_4) A_{2c}^{(j)} (-t_3), \label{impifin}
\eeq
where $w_{1,2} = (z \pm y)^2 \pm (z \pm y)$ and  $w_{3,4}  = -(z \pm y)^2 
\pm (z \pm y)$. The functions $A_{2c}^{(j)}$ and the arguments $t_1$ and 
$t_3$ are defined in Eqs. \eq{a2c}, \eq{t12} and \eq{t34}, respectively. 
On the analogy of $A_{2c}^{(1)}$ explicitly quoted in Ref.~\onlinecite{pletgr},
one can as well express the rest $A_{2c}^{(j)}$ in terms of elliptic functions.
For example, in Eqs.~\eq{a0ye} and \eq{azze} we write down explicit formulas 
for $A_{2c}^{(2)}$ and  $A_{2c}^{(4)}$, respectively.

\begin{figure}[t]
\includegraphics[width =8.25cm]{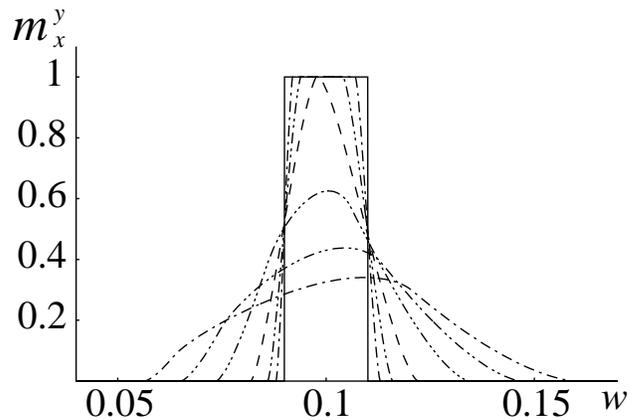}
\caption{Spin-galvanic function \eq{mnxy} at $y=0.1$ and the values of $z=0.0$,
$0.025$, $0.005$, $0.01$, $0.02$, $0.03$,  $0.04$. In particular, the solid 
line depicts the long-wavelength limit, and the dashed curve corresponds to 
$z=y^2 =0.01 \approx z^*$. }
\label{galvan}
\end{figure}

Eq. \eq{impifin} allows us to study the behavior of the density-density 
response functions at finite momenta. On its basis we can also describe the 
spin-density  response $\rho^{y} = 2 M^y_x E_x$ to a longitudinal electric 
field $E_x = -e  \frac{\partial V^0}{\partial x}$ (or $E_x = -i q V^0$ in the 
momentum representation). The spin-galvanic response 
function\cite{edelst,rasef,shekhter} $M^y_x$ is then related to $\chi^{0y}$ 
via $M_x^y = i e \chi^{0y}/(2 q)$.
Let us also introduce the rescaled function
\be
m^y_x = -\frac{16 k_R}{e \pi \nu} {\rm Re} M^y_x  \equiv \frac{4 y}{\pi z \nu} 
{\rm Im} \chi^{0y}.
\label{mnxy}
\ee
Using \eq{a0ye}, we calculate and plot it in Fig.~\ref{galvan} at fixed finite 
values of $z$. We remark that its frequency dependence in the range 
$w_4 < w < w_1$ corresponding to intersubband transitions is similar to that 
of the finite-$q$ conductivity studied in Ref.~\onlinecite{pletgr}.

\subsection{Long-wavelength limit}
\label{longapp}

Spatially uniform spin susceptibilities of the 2DEG with SO coupling have been 
previously considered  in Ref.~\onlinecite{schliem}. We recover the 
corresponding expressions in the long-wavelength limit
\beq
 - \lim_{z \to 0} \frac{\chi^{zz}}{\nu} = - 2 \lim_{z \to 0} 
\frac{\chi^{xx (yy)}}{\nu} = 2 + \frac{w}{2 y^2} r (w) ,
\eeq
where
\beq
r (w) = \ln \bigg|\frac{(w-y^2)^2- y^2 }{(w+y^2)^2 -y^2} \bigg| 
+ i \pi \Theta (y^2 -|w-y|).
\label{rw}
\eeq

The small-$q$ behavior of the polarization operator of 2DEG with Rashba SO 
coupling has been approximated in Refs.~\onlinecite{MH,MCE} by the expression
\beq
-\frac{\chi^{00}}{\nu} \approx -\frac{z^2}{w^2} +\frac{z^2}{4 w} r (w),
\label{polnul}
\eeq
which is obtained after the formal expansion of $\chi^{00}$ in a series of 
$z  = \frac{q}{2 k_F} \ll 1$. Later on it has been remarked\cite{pletgr} that 
the formula \eq{polnul} is, in fact, reliable only for the values $z < z^* 
\approx y^2$, where $z^*$ is the point of intersection of the parabolas 
$w_2 (z)$ and $w_3 (z)$ [see Fig.~\ref{zstar}]. Analogously, we can find 
approximate relations for the off-diagonal terms
\beq
- \frac{i \chi^{zx}}{\nu} \approx  \frac{z w}{2 y^3} \left[ r (w) 
+ \sum_{\mu=\pm} \frac{w y (\mu - 2 y)}{y^2 (1 -\mu y)^2 -w^2} \right] 
\eeq
and
\beq
\frac{\chi^{0y}}{\nu} & \approx &  \frac{z}{4 y} \left[ r (w) 
+ \frac{4 y^2}{w}\right],
\label{h0yl}
\eeq
which are also applicable in the small triangular region located at $z<z^*$ 
and bounded by $w_3 <w <w_2$.  In particular, Eq.~\eq{h0yl} accounts for the 
box-like shape of the function $m_x^y$ \eq{mnxy} in the linit $z \to 0$ which 
can be seen in Fig.~\ref{galvan}.

\subsection{Quasiclassical approximation}
\label{quasicl}

On the basis of the exact result \eq{impifin} we can also estimate how 
accurate the quasiclassical approximation appears to be, when it is applied 
to a description of the clean 2DEG with Rashba SO coupling.

The quasiclassical approximation relies on the fact that all energy scales in 
the system are much smaller than the Fermi energy: $q k_F/m^*, \omega, 
\alpha_R k_F \ll \epsilon_F$. This inequality enables one to treat 
\eq{pilm00}-\eq{pilm0y} linearizing  the branches of the spectrum 
$\epsilon^{\mu}_{{\bf k}}$ near the corresponding Fermi points 
$k_{\mu}=k_F -\mu k_R$ and expanding  ${\cal F}^{\alpha \beta}_{{\bf k}, 
{\bf k}+{\bf q}; \mu, \mu'}$ in a series of $q$. We note that a more 
systematic procedure of making the quasiclassical approximation is based 
on the gradient expansion of the quantum kinetic equation\cite{MH,MSH,malsh}, 
which is applicable in more general -- nonequilibrium -- situations.

Quasiclassical response functions $\chi^{\alpha \beta}$ of the disordered 2DEG 
with Rashba SO coupling have been explicitly calculated in 
Ref.~\onlinecite{pletcr}. It has been also argued therein that the 
quasiclassical approximation is justified in the presence of  a finite 
amount of impurities such that $\tau^{-1} >k_R^2/m^*$. This condition 
confines $\tau^{-1}$ from below. Had we ignored it, we would have obtained 
in the extreme collisionless limit $\tau \to \infty$ the quasiclassical 
expression for, say, out-of-plane component of the spin susceptibility in 
the form 
\beq
- \frac{1}{\nu} {\rm Im} \chi^{zz}_{qc} = w \sum_{\mu = \pm} 
\frac{\Theta (z^2-(w -\mu y)^2)}{\sqrt{z^2 - (w -\mu y)^2}}, 
\label{qczz}
\eeq
which is divergent near $w = w^{qc}_{\pm} \equiv |y\pm z|$. 

The lines  $w^{qc}_{\pm}$ represent the quasiclassical boundaries of the 
SO-induced (intersubband) particle-hole excitation region, and they differ 
from the actual parabolic ones $w_+  \equiv w_1$ and $w_- \equiv w_4$. In 
Fig.~\ref{zstar} $ w^{qc}_{\pm}$ are depicted by the dashed lines, and one 
can observe that $w^{qc}_+$ lies in between $w_1$ and $w_3$, and  $w^{qc}_-$ 
lies in between $w_2$ and $w_4$. It is also evident that in the quasiclassical 
approximation the finite basis $y-y^2 <w <y+y^2$ of the wedge  at $z =0$ is 
not resolved. This means that the imaginary part of a quasiclassical 
counterpart of $r (w)$ \eq{rw} appears to  be delta-peaked, and therefore  
the corresponding finite-valued results of Sec.~\ref{longapp} can not be 
reproduced in the quasiclassical approximation. 

\begin{figure}[t]
\includegraphics[width =8.25cm]{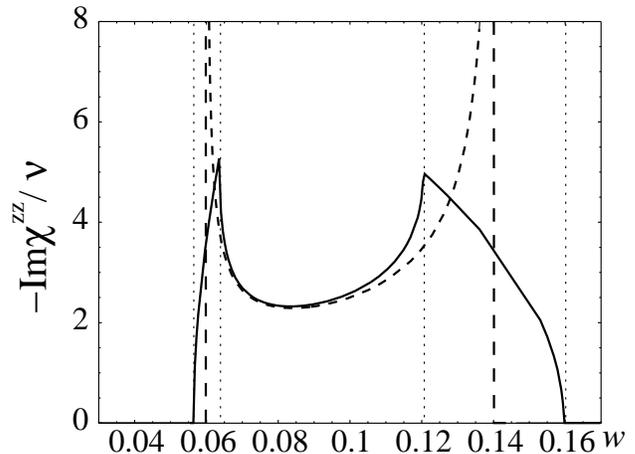}
\caption{Comparison of the exact (solid line) and the quasiclassical 
(dashed line) functions $-{\rm Im} \chi^{zz}/\nu$ at $z=0.04$ and $y=0.1$. 
Vertical dotted lines correspond (from the left to the right) to the frequency 
values $w_4$, $w_2$, $w_3$, and $w_1$. Vertical dashed lines correspond to the 
values $w_-^{qc}=y-z$ and $w_+^{qc}=y+z$ demarcating the quasiclassical 
boundaries of the SO-induced particle-hole excitation region.}
\label{mzzqcl}
\end{figure}

It can be anticipated that the quasiclassical approximation is still reliable 
in the triangular area $w_2 < w<w_3$ at $z>z^* \approx y^2$. In 
Fig.~\ref{mzzqcl} we confirm this surmise comparing the function \eq{qczz} 
with the exact expression calculated on the basis of Eqs.~\eq{impifin} and 
\eq{azze}. 

Thus, we conclude that the long-wavelength and the quasiclassical 
approximations do not have any correspondence between each other, since they 
are applicable in the domains which do not overlap, i.e. at $z<z^*$ and 
$z>z^*$, respectively. 
 
\section{Collective charge and spin density excitations}
\label{collect}

Let us now consider the renormalization of the matrix $\chi$ due to 
electron-electron interaction. Treating the latter in the Hubbard 
approximation\cite{Mahan}, we take into account screening and exchange effects.
An effective response matrix $\widetilde{\chi}$ [see Eq. \eq{renchi}] is 
derived in Appendix \ref{eom} using the method of the equations of 
motion\cite{Mahan}.

In the basis specified by the condition ${\bf q} || {\bf e}_x$ the matrix 
$\chi$ can be decomposed into two $2\times2$-blocks
\be
\chi_{(0y)} = \left( \begin{array}{cc} \chi^{00} & \chi^{0y} \\ \chi^{y0} & 
\chi^{yy} \end{array} \right), \quad \chi_{(xz)} = \left( \begin{array}{cc} 
\chi^{xx} & \chi^{xz} \\ \chi^{zx} & \chi^{zz} \end{array} \right).
\ee 
It follows from \eq{renchi} that the spin-charge $\chi_{(0y)}$ and the 
spin-spin $\chi_{(xz)}$ blocks are renormalized independently of each other. 
In particular,
\beq
\widetilde{\chi}_{(0y)} &=& \left(1-\chi_{(0y)} F_{(0y)} \right)^{-1} 
\chi_{(0y)}, \label{col0y} \\
\widetilde{\chi}_{(xz)} &=& \left(1 + J \chi_{(xz)} \right)^{-1} \chi_{(xz)}, 
\label{colxz}
\eeq
where $F_{(0y)} = {\rm diag} \{ v_q -J, -J \} \approx {\rm diag} 
\{ v_q, -J \}$, $v_q = \frac{2 \pi e^2}{q}$ is the Coulomb interaction, and 
$J = \frac{\pi e^2}{k_F}$ is the Hubbard's vertex exchange term at small 
$q \ll k_F$. 

Let us first consider the spin-charge block and find the renormalized response 
functions
\beq
\widetilde{\chi}^{00} &=& \frac{\chi^{00} 
+ J \bar{\Delta}_{(0y)}}{\Delta_{(0y)}}, \label{wt00} \\
\widetilde{\chi}^{yy} &=& \frac{\chi^{yy} -v_q 
\bar{\Delta}_{(0y)}}{\Delta_{(0y)}}, \label{wtyy} \\
\widetilde{\chi}^{0y} &=& \frac{\chi^{0y}}{\Delta_{(0y)}}, 
\label{wt0y}
\eeq
where
\beq
\bar{\Delta}_{(0y)} &=& \det \chi_{(0y)} = \chi^{00} \chi^{yy} - 
\left( \chi^{0y} \right)^2 ,\\
\Delta_{(0y)} &=& \det \left(1- \chi_{(0y)} F_{(0y)}\right) \nonumber \\
& \approx & \left( 1 - v_q \chi^{00} \right) \left( 1+ J \chi^{yy} \right)+ 
J v_q\left( \chi^{0y} \right)^2  .
\eeq
For the spin-spin block we have
\beq
\widetilde{\chi}^{xx} &=& \frac{\chi^{xx} 
+ J \bar{\Delta}_{(zx)}}{\Delta_{(zx)}}, \label{wtxx} \\
\widetilde{\chi}^{zz} &=& \frac{\chi^{zz} 
+J \bar{\Delta}_{(zx)}}{\Delta_{(zx)}}, \label{wtzz} \\
\widetilde{\chi}^{xz} &=& \frac{\chi^{xz}}{\Delta_{(xz)}}, 
\label{wtxz}
\eeq
where
\beq
\bar{\Delta}_{(xz)} &=& \det \chi_{(zx)} = \chi^{xx} \chi^{zz} 
+ \left( \chi^{xz} \right)^2 ,\\
\Delta_{(xz)} &=& \det \left(1+J \chi_{(xz)}\right) \nonumber \\
&=& \left( 1 +J \chi^{xx} \right) \left( 1+ J \chi^{zz} \right)
+ J^2 \left( \chi^{xz} \right)^2  .
\eeq

Dispersions of the collective charge and spin density excitations are 
determined from the equations $\Delta_{(0y)}=0$ and $\Delta_{(xz)}=0$, 
and the response functions $\widetilde{\chi}^{\alpha \beta}$ are strongly 
enhanced at the parameter values satisfying these conditions.

Before quantifying $\widetilde{\chi}^{\alpha \beta}$, let us qualitatively 
discuss the role of the spin-charge mixing term $\chi^{0y}$ as well as the 
role of the exchange corrections. Note that if either $\chi^{0y}=0$ or $J=0$, 
the dynamics of the charge and the spin-$y$ densities becomes decoupled, and
 the corresponding plasmon and SDE$_y$ (spin-$y$ density excitation) modes 
are independently found from the conditions $1 -v_q \chi^{00}=0$ and 
$1+J \chi^{yy}=0$ (the latter equation makes sense at $J \neq 0$ only). 
The role of $\chi^{0y}$, whatever small it might be, is considerably 
strengthened when  the dispersions of the plasmon and the  SDE$_y$ modes 
come close to a degeneracy point. In fact, $\chi^{0y} \neq 0$ lifts this 
degeneracy, thus making a possible crossing of these modes avoidable. 
Therefore, we expect that an  account of the spin-charge mixing along 
with the exchange interaction would lead to nontrivial features in 
profiles of $\widetilde{\chi}^{00}$ and $\widetilde{\chi}^{yy}$ near 
the avoided intersection. It is also implied that the peaks corresponding 
to the both collective modes would appear in every response function of 
the spin-charge block. In particular, the spin susceptibility 
$\widetilde{\chi}^{yy}$ is expected to manifest  resonant features at the 
position of the plasmon dispersion, while $\widetilde{\chi}^{00}$ should 
have a peak corresponding to the  SDE$_y$ mode.
     
We remark that the decoupled plasmon mode in 2DEG with Rashba SO coupling 
has been previously considered  in 
Refs.~\onlinecite{xu,wang,gumbs,kush1,pletgr} in the random phase 
approximation (RPA), i.e. at $J=0$. In turn, the coupled spin-$x$ -- spin-$z$ 
collective modes, which are determined by $\Delta_{(xz)}=0$, occur at 
$J \neq 0$, i.e. their description requires an extension of  the RPA. Their 
dispersions, as well as the dispersion of the decoupled  spin-$y$ mode 
(at neglected spin-charge mixing $\chi^{0y}=0$), have been previously 
considered in Ref.~\onlinecite{brataas} in terms of the Hubbard's 
approximation  with the bare spin susceptibilities calculated in the 
quasiclassical approximation (see Sec.~\ref{quasicl}).
 
\begin{figure}[t]
\includegraphics[width =8.25cm]{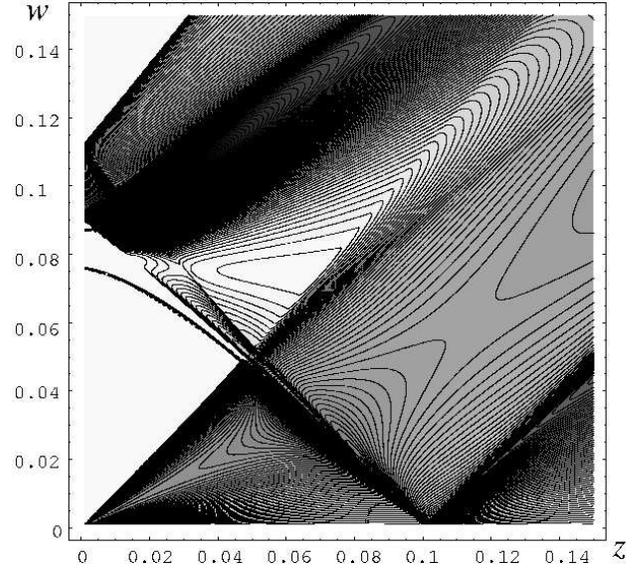}
\caption{Contour plot of $-{\rm Im} \widetilde{\chi}^{xx} (z,w)/\nu$. 
Parameters: $y=0.1$ and $r_s =0.6$.}
\label{mxxcp}
\end{figure}

Using our exact expressions for $\chi^{\alpha \beta}$, we are able to study 
dispersions of the collective charge and spin density  modes in more detail. 
They can be visualized, for example, in contour plots of the response 
functions $\widetilde{\chi}^{\alpha \beta}$.  Absolute values of  
$\widetilde{\chi}^{\alpha \beta}$ providing a useful information for 
inelastic Raman scattering\cite{raman} are available as well.

Let us start our consideration from the renormalized response functions 
\eq{wtxx}-\eq{wtxz} constituting the spin-spin block $\widetilde{\chi}_{(xz)}$.
 In Fig.~\ref{mxxcp} we present the contour plot of $- {\rm Im} 
\widetilde{\chi}^{xx}/\nu$ at $y=0.1$ and the Wigner-Seitz parameter 
$r_s\equiv \frac{\sqrt{2} m^* e^2}{k_F} =0.6$ (note that $J=\frac{r_s}{2 
\sqrt{2} \nu}$). One can observe the dispersions of the two coupled spin-$x$ 
-- spin-$z$ collective modes. We state that their spectra are in the 
qualitative agreement with those previously predicted in 
Ref.~\onlinecite{brataas} on the basis of the quasiclassical approximation. 
However, the absolute values of $\widetilde{\chi}^{\alpha \beta}$ differ 
from their quasiclassical counterparts $\widetilde{\chi}_{qc}^{\alpha \beta}$. 
The reason is the same as discussed in Sec.~\ref{quasicl} for the case of bare 
susceptiblities.

\begin{figure}[t]
\includegraphics[width =8.25cm]{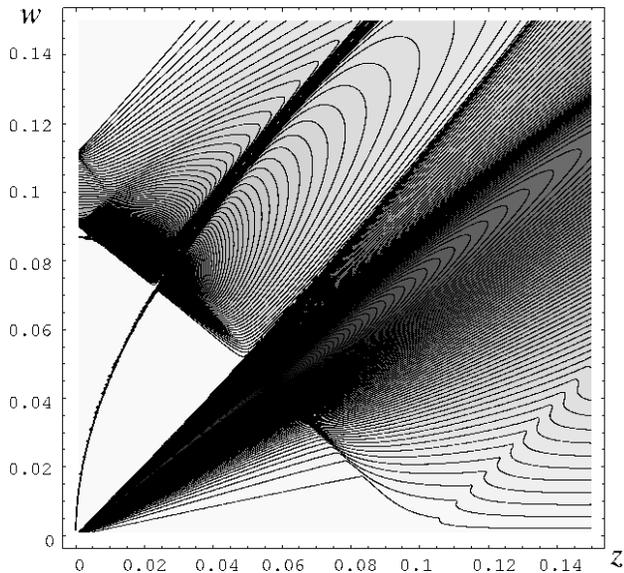}
\caption{Contour plot of $-{\rm Im} \widetilde{\chi}^{yy} (z,w)/\nu$. 
Parameters: $y=0.1$ and $r_s =0.6$. }
\label{myycp}
\end{figure}

\begin{figure}[t]
\includegraphics[width =8.25cm]{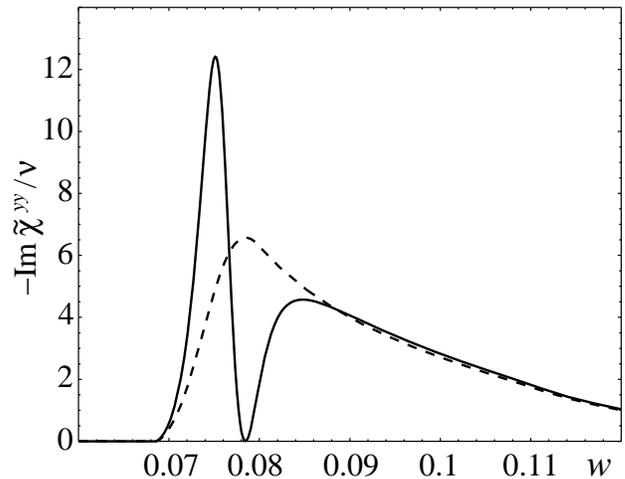}
\caption{The cross section of $-{\rm Im} \widetilde{\chi}^{yy} (z,w)/\nu$ at 
$z=0.027$ (solid line). The dashed line corresponds to the case of 
intentionally neglected spin-charge mixing term $\chi^{0y}$. 
Parameters: $y=0.1$ and $r_s=0.6$.}
\label{myysec}
\end{figure}

In turn, the response functions \eq{wt00}-\eq{wt0y} of the spin-charge block 
$\widetilde{\chi}_{(0y)}$ manifest novel qualitative features due to the 
account of the spin-charge mixing $\chi^{0y}$ along with the exchange 
interaction.
In Fig.~\ref{myycp} we present the contour plot of the function $- {\rm Im} 
\widetilde{\chi}^{yy}/\nu$ at $y=0.1$ and $r_s =0.6$ (note that $v_q = 
\frac{r_s}{2 \sqrt{2} z \nu}$). It also contains the  two collective modes: 
plasmon-like and SDE$_y$-like. The plasmon-like mode has almost the same  
dispersion $w \approx \sqrt{\frac{r_s z}{2 \sqrt{2}}}$ as the plasmon mode in 
the absence of SO coupling. The SDE$_y$-like mode originates at $z=0$ from the 
finite frequency value slightly below the bottom of the wedge. Being undamped, 
these modes do not come close to each other. They collide soon after the 
plasmon-like mode enters into the  SO-induced damping region, i.e. at 
$z \geq 0.023$. In the vicinity of this point the spin-charge mixing 
$\chi^{0y}$ acquires its importance. Although a pictorial description 
of the avoided crossing loses its obviousness because of the modes' 
broadening, we prove that it does happen. For this purpose we plot in 
Fig.~\ref{myysec} the cross-section of $- {\rm Im} \widetilde{\chi}^{yy}/\nu$ 
at $z=0.027$. Thereby we show that instead of the only SDE$_y$ peak 
(dashed line) occurring at intentionally neglected $\chi^{0y}=0$, we 
obtain at $\chi^{0y} \neq 0$ the two well-resolved peaks (solid line) 
corresponding to the plasmon-like and the SDE$_y$-like modes. A formation 
of the dip between them is interpreted as an avoided crossing of the two 
broadened modes. A relevance of such interpretation becomes even more 
evident, if one would compare several subsequent cross sections of 
$- {\rm Im} \widetilde{\chi}^{yy}/\nu$ and  $- {\rm Im} 
\widetilde{\chi}^{00}/\nu$.

\begin{figure}[b]
\includegraphics[width =8.25cm]{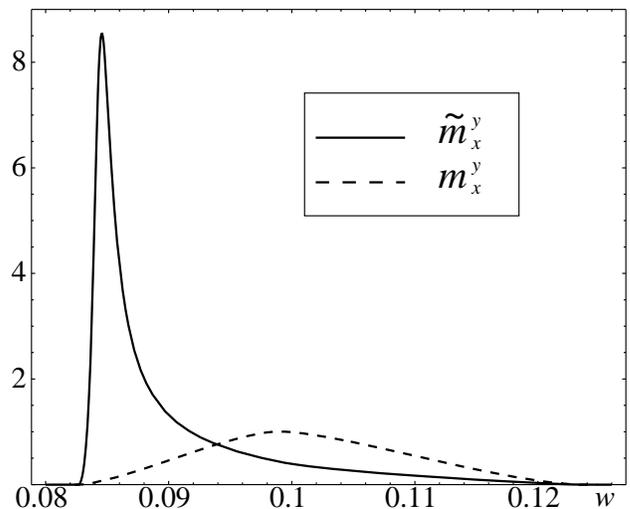}
\caption{The renormalized spin-galvanic function \eq{mrxy} at $z=0.01$ and 
the parameters $y=0.1$ and $r_s=0.6$ (solid line). The dashed line depicts 
its non-interacting counterpart ($r_s =0$) given by \eq{mnxy}.}
\label{ren0y}
\end{figure}

As one can see from Eq.~\eq{wt0y}, the spin-charge mixing term is renormalized 
by electron-electron interaction as well. Let us define the function
\be
\widetilde{m}_x^y = \frac{4 y}{\pi z \nu} {\rm Im} \widetilde{\chi}^{0y},
\label{mrxy}
\ee  
which is an interacting counterpart of the bare spin-galvanic function $m_x^y$ 
\eq{mnxy}. In Fig.~\ref{ren0y} we compare both of them at $z=0.01$ and $y=0.1$ 
in the frequency window $y - 2 y^2 < w < y+ 2 y^2$. We  observe that at these 
parameters the major effect of renormalization consists in a considerable 
amplification of the spin-galvanic function \eq{mrxy} due to the exchange 
interaction.

\section{Summary}

We have derived zero-temperature analytic expressions for the charge and spin 
density response functions of the clean 2DEG with Rashba SO coupling at finite 
momenta and frequencies. We have studied their static and long-wavelength 
limits as well as established the applicability range of the quasiclassical 
approximation. In the static limit we have observed the Kohn-like anomalies 
in the spin susceptibilities and showed how they are related to the SO 
modification of the RKKY interaction.

The renormalization of the response functions due to electron-electron 
interaction has been considered in the Hubbard's approximation. We have 
studied the collective charge and spin density modes which appear to be 
coupled due to the nonvanishing spin-charge mixing term $\chi^{0y}$. One of 
the important consequences of this coupling is the emergence of the 
plasmon-like peak in the spectrum of the renormalized spin susceptibility.

\section{Acknowledgements}
We are grateful to Gerd Sch\"on for useful discussions. M.P. acknowledges the  
financial support from the Deutsche Forschungsgemeinschaft (DFG).

\appendix

\section{Equations of motion}
\label{eom}

Let us define the local charge ($\alpha=0$) and spin ($\alpha=x,y,z$) density 
operators
\be
\hat{\rho}^{\alpha}_{{\bf q}} = \sum_{{\bf k}} \hat{\rho}^{\alpha}_{{\bf k} 
{\bf q}}, \quad
\hat{\rho}^{\alpha}_{{\bf k} {\bf q}} = c^{\dagger}_{{\bf k}} \sigma^{\alpha}
c_{{\bf k}+{\bf q}},
\ee
and derive the equations of motion\cite{Mahan} for the expectation values  
$\rho^{\alpha}_{{\bf k} {\bf q}} =\langle \hat{\rho}^{\alpha}_{{\bf k} 
{\bf q}}\rangle$. In the Fourier representation they read
\be
\omega \rho^{\alpha}_{{\bf k} {\bf q}} = \langle [\hat{\rho}^{\alpha}_{{\bf k} 
{\bf q}}, H ] \rangle .
\ee
The full Hamiltonian $H = H_0 + H_{{\rm ext}}$ consists of the parts 
describing a system 
\be
H_{0} = \sum_{{\bf k}',\beta} E_{{\bf k}'}^{\beta} 
\hat{\rho}^{\beta}_{{\bf k}',0}
\ee
and an external perturbation
\be
H_{{\rm ext}} = \frac{1}{{\cal V}} \sum_{{\bf q}',\beta} V_{{\bf q}'}^{\beta} 
\hat{\rho}^{\beta}_{- {\bf q}'},
\label{hext}
\ee
where ${\cal V}$ is the spatial volume. For the Rashba system \eq{initham} we 
have $E_{{\bf k}}^{\beta} = (\epsilon_{{\bf k}}, 
\alpha_R k_y, -\alpha_R k_x ,0)$. Later on we will also add the interaction 
term $H_{{\rm int}}$ to the Hamiltonian $H$.

Using the identity
\beq
[ \hat{\rho}^{\alpha}_{{\bf k} {\bf q}}, \hat{\rho}^{\beta}_{{\bf k}',0} ] = 
c_{{\bf k}}^{\dagger} \sigma^{\alpha} \sigma^{\beta} c_{{\bf k}'} 
\delta_{{\bf k}+{\bf q},{\bf k}'} -c_{{\bf k}'}^{\dagger} \sigma^{\beta} 
\sigma^{\alpha} c_{{\bf k}+{\bf q}} \delta_{{\bf k} {\bf k}'}
\eeq
we find
\beq
\langle [ \hat{\rho}^{\alpha}_{{\bf k} {\bf q}},  H_0] \rangle &=& 
\sum_{\beta} E_{{\bf k}+{\bf q}}^{\beta} \langle c_{{\bf k}}^{\dagger} 
\sigma^{\alpha} \sigma^{\beta} c_{{\bf k}+{\bf q}}\rangle \nonumber \\
&-& \sum_{\beta} E_{{\bf k}}^{\beta} \langle c_{{\bf k}}^{\dagger} 
\sigma^{\beta} \sigma^{\alpha} c_{{\bf k}+{\bf q}} \rangle .
\eeq
It is convenient to introduce the representation
\be
\omega \rho^{\alpha}_{{\bf k} {\bf q}} - \langle [ 
\hat{\rho}^{\alpha}_{{\bf k} {\bf q}},  H_0] \rangle = \sum_{\beta} 
{\cal A}^{\alpha \beta}_{{\bf k} {\bf q}}\rho^{\beta}_{{\bf k} {\bf q}}
\label{adef}
\ee
in terms of the matrix ${\cal A}$ with the elements
\beq
{\cal A}^{\alpha \alpha}_{{\bf k} {\bf q}} &=& \omega +\epsilon_{{\bf k}} 
- \epsilon_{{\bf k}+{\bf q}},\nonumber \\
{\cal A}^{x0}_{{\bf k} {\bf q}} &=& {\cal A}^{0x}_{{\bf k} {\bf q}} 
=-\alpha_R q_y,\nonumber \\
{\cal A}^{y0}_{{\bf k} {\bf q}} &=& {\cal A}^{0y}_{{\bf k} {\bf q}} 
=\alpha_R q_x,\nonumber \\
{\cal A}^{xz}_{{\bf k} {\bf q}} &=& -{\cal A}^{zx}_{{\bf k} {\bf q}} 
=i \alpha_R (2 k_x + q_x ),\nonumber \\
{\cal A}^{yz}_{{\bf k} {\bf q}} &=& -{\cal A}^{zy}_{{\bf k} {\bf q}} 
=i \alpha_R (2 k_y + q_y ),\nonumber \\
{\cal A}^{0z}_{{\bf k} {\bf q}} &=& {\cal A}^{z0}_{{\bf k} {\bf q}} 
={\cal A}^{xy}_{{\bf k} {\bf q}} = {\cal A}^{yx}_{{\bf k} {\bf q}} =0 .
\eeq

In order to find the commutator of $\hat{\rho}^{\alpha}_{{\bf k} {\bf q}}$ 
with $H_{{\rm ext}}$ \eq{hext}, we use the identity
\beq
[ \hat{\rho}^{\alpha}_{{\bf k} {\bf q}}, \hat{\rho}^{\beta}_{- {\bf q}'} ] = 
c_{{\bf k}}^{\dagger} \sigma^{\alpha} \sigma^{\beta} c_{{\bf k}+{\bf q}
-{\bf q}'} -c_{{\bf k}+{\bf q}'}^{\dagger} \sigma^{\beta} \sigma^{\alpha} 
c_{{\bf k}+{\bf q}}.
\eeq
From the whole sum over ${\bf q}'$ in \eq{hext} we pick out the only term 
${\bf q}'={\bf q}$. This is known as the random phase approximation (RPA). 
Then,
\be
[ \hat{\rho}^{\alpha}_{{\bf k} {\bf q}},  H_{{\rm ext}}] \approx  
\frac{1}{{\cal V}} \sum_{\beta} V_{{\bf q}}^{\beta}\left(c_{{\bf k}}^{\dagger} 
\sigma^{\alpha} \sigma^{\beta} c_{{\bf k}} -c_{{\bf k}+{\bf q}}^{\dagger} 
\sigma^{\beta} \sigma^{\alpha} c_{{\bf k}+{\bf q}} \right),
\ee
and after averaging we obtain 
\be
\langle [ \hat{\rho}^{\alpha}_{{\bf k} {\bf q}},  H_{{\rm ext}}] \rangle 
\approx \frac{1}{{\cal V}} \sum_{\beta}{\cal B}^{\alpha \beta}_{{\bf k} 
{\bf q}} V_{{\bf q}}^{\beta},
\label{bdef}
\ee 
where 
\be
{\cal B}^{\alpha \beta}_{{\bf k} {\bf q}} = \langle c_{{\bf k}}^{\dagger} 
\sigma^{\alpha} \sigma^{\beta} c_{{\bf k}} -c_{{\bf k}+{\bf q}}^{\dagger} 
\sigma^{\beta} \sigma^{\alpha} c_{{\bf k}+{\bf q}} \rangle.
\label{be1}
\ee

In order to fulfil the averaging, we have to transform $c_{{\bf k}} 
= {\cal U}_{{\bf k}} \gamma_{{\bf k}}$ into the diagonal basis 
$\gamma_{{\bf k}\pm}$ such that 
\be
\langle \gamma_{{\bf k} \mu}^{\dagger} \gamma_{{\bf k} \mu'} \rangle 
= \delta_{\mu \mu'} f_{{\bf k}}^{\mu}.
\label{gav}
\ee
The transformation  matrix ${\cal U}_{{\bf k}}$ is defined in Eq.~\eq{ukm}. 
Since the external perturbation is assumed to be small, we average in \eq{gav} 
with respect to the system's density matrix. Therefore we can identify 
$f_{{\bf k}}^{\mu} \equiv n_F (\epsilon_{{\bf k}}^{\mu})$.

The components \eq{be1} are then found to be
\beq
{\cal B}^{\alpha \beta}_{{\bf k} {\bf q}} &=& \frac12 \sum_{\mu} 
f_{{\bf k}}^{\mu} {\rm Tr} \left[ (1+\mu \sigma^z) 
{\cal U}_{{\bf k}}^{\dagger}\sigma^{\alpha} \sigma^{\beta} {\cal U}_{{\bf k}}
\right] \nonumber \\
&-& \frac12 \sum_{\mu} f_{{\bf k}+{\bf q}}^{\mu} {\rm Tr} \left[ (1+\mu 
\sigma^z) {\cal U}_{{\bf k}+{\bf q}}^{\dagger}\sigma^{\beta} \sigma^{\alpha} 
{\cal U}_{{\bf k}+{\bf q}}\right] \nonumber \\
&=& \frac12 \sum_{\mu} f_{{\bf k}}^{\mu} {\rm Tr} \left[ (1+\mu h^R_{{\bf k}}) 
\sigma^{\alpha} \sigma^{\beta} \right] \nonumber \\
&-& \frac12 \sum_{\mu} f_{{\bf k}+{\bf q}}^{\mu} {\rm Tr} \left[ 
(1+\mu h^R_{{\bf k}+{\bf q}}) \sigma^{\beta} \sigma^{\alpha} \right],
\eeq
or, more explicitly,
\beq
{\cal B}^{\alpha \alpha}_{{\bf k} {\bf q}} &=& \sum_{\mu} 
\left(f_{{\bf k}}^{\mu} - f_{{\bf k}+{\bf q}}^{\mu} \right) , \nonumber \\ 
{\cal B}^{x0}_{{\bf k} {\bf q}} &=& {\cal B}^{0x}_{{\bf k} {\bf q}} 
= \sum_{\mu} \mu \left(f_{{\bf k}}^{\mu} \sin \phi_{{\bf k}}
- f_{{\bf k}+{\bf q}}^{\mu} \sin \phi_{{\bf k}+{\bf q}} \right), \nonumber \\
{\cal B}^{y0}_{{\bf k} {\bf q}} &=& {\cal B}^{0y}_{{\bf k} {\bf q}} 
= -\sum_{\mu} \mu \left(f_{{\bf k}}^{\mu} \cos \phi_{{\bf k}}
- f_{{\bf k}+{\bf q}}^{\mu} \cos \phi_{{\bf k}+{\bf q}} \right), \nonumber \\
{\cal B}^{xz}_{{\bf k} {\bf q}} &=& -{\cal B}^{zx}_{{\bf k} {\bf q}} 
= i \sum_{\mu} \mu \left(f_{{\bf k}}^{\mu} \cos \phi_{{\bf k}}
+ f_{{\bf k}+{\bf q}}^{\mu} \cos \phi_{{\bf k}+{\bf q}} \right), \nonumber \\
{\cal B}^{yz}_{{\bf k} {\bf q}} &=& -{\cal B}^{zy}_{{\bf k} {\bf q}} 
= i \sum_{\mu} \mu \left(f_{{\bf k}}^{\mu} \sin \phi_{{\bf k}}
+ f_{{\bf k}+{\bf q}}^{\mu} \sin \phi_{{\bf k}+{\bf q}} \right),\nonumber \\
{\cal B}^{0z}_{{\bf k} {\bf q}} &=& {\cal B}^{z0}_{{\bf k} {\bf q}} 
={\cal B}^{xy}_{{\bf k} {\bf q}} = {\cal B}^{yx}_{{\bf k} {\bf q}} =0 .
\eeq

Combining \eq{adef} and \eq{bdef}, we derive the following matrix equation
\be
{\cal A}_{{\bf k} {\bf q}} \rho_{{\bf k} {\bf q}} =  \frac{1}{{\cal V}}
{\cal B}_{{\bf k} {\bf q}} V_{{\bf q}},
\label{ebare}
\ee
where the upper indices are omitted for brevity. Inverting the matrix 
${\cal A}_{{\bf k} {\bf q}}$ and summing over ${\bf k}$, we obtain 
$\rho_{{\bf q}} \equiv \langle \hat{\rho}_{{\bf q}} \rangle = \chi V_{\bf q}$, 
where 
\be
\chi = \frac{1}{\cal V} \sum_{{\bf k}} {\cal A}_{{\bf k} {\bf q}}^{-1} 
{\cal B}_{{\bf k} {\bf q}}
\label{cbare}
\ee 
is a density response matrix of the non-interacting system. After the 
straightforward calculation, we recover from \eq{cbare} the expression 
\eq{chidef} for the components of $\chi$. 

Let us now take into account  electron-electron interaction
\be
H_{{\rm int}} = \frac{1}{2 {\cal V}} \sum_{{\bf p}' {\bf k}' {\bf q}'} 
\sum_{\sigma s} v_{{\bf q}'} c^{\dagger}_{{\bf p}'+{\bf q}',\sigma} 
c^{\dagger}_{{\bf k}'-{\bf q}',s} c_{{\bf k}' s} c_{{\bf p}' \sigma}.
\ee
In the mean field approximation we obtain
\beq
& & \langle [ \hat{\rho}^{\alpha}_{{\bf k} {\bf q}}, H_{{\rm int}}] \rangle
= \frac{1}{{\cal V}} \sum_{{\bf q}'} v_{{\bf q}'} \rho_{{\bf q}'}^0 \times 
\nonumber \\
& & \qquad \times \left\{  \langle c_{{\bf k}}^{\dagger} \sigma^{\alpha} 
c_{{\bf k}+{\bf q}-{\bf q}'}  \rangle - \langle c^{\dagger}_{{\bf k}+{\bf q}'} 
\sigma^{\alpha} c_{{\bf k}+{\bf q}}  \rangle \right\} \nonumber \\
&+& \frac{1}{2 {\cal V}} \sum_{{\bf q}',{\bf p}',\beta} v_{{\bf q}'} 
\left\{  \langle c^{\dagger}_{{\bf k}+{\bf q}'} \sigma^{\beta} c_{{\bf p}'}  
\rangle \langle c^{\dagger}_{{\bf p}'-{\bf q}'} \sigma^{\beta} \sigma^{\alpha} 
c_{{\bf k}+{\bf q}} \rangle \right. \nonumber \\
& & \qquad \quad \left. - \langle  c_{{\bf k}}^{\dagger} \sigma^{\alpha} 
\sigma^{\beta} c_{{\bf p}'} \rangle \langle c^{\dagger}_{{\bf p}'+{\bf q}'}  
\sigma^{\beta} c_{{\bf k}+{\bf q}+{\bf q}'} \rangle   \right\} .
\label{eegen}
\eeq

The first sum  in \eq{eegen} corresponds to the direct Coulomb term. We treat 
it further in the RPA picking the only term ${\bf q}'={\bf q}$ out of the 
whole sum. We obtain the following contribution
\beq
 & & \langle [ \hat{\rho}^{\alpha}_{{\bf k} {\bf q}}, H_{{\rm int}}] 
\rangle^{{\rm RPA}} = \label{scr} \\
& & = \frac{v_{{\bf q}} \rho_{{\bf q}}^0}{{\cal V}}  \left\{  
\langle c_{{\bf k}}^{\dagger} \sigma^{\alpha} c_{{\bf k}}  \rangle 
-  \langle c^{\dagger}_{{\bf k}+{\bf q}} \sigma^{\alpha} c_{{\bf k}+{\bf q}}  
\rangle \right\} = \frac{v_{{\bf q}} \rho_{{\bf q}}^0}{{\cal V}} 
{\cal B}^{\alpha 0}_{{\bf k} {\bf q}}, 
\nonumber
\eeq 
which accounts for the effect of screening.

From the second sum in \eq{eegen} we can extract the exchange self-energy term 
and the Hubbard's exchange correction to the RPA. 

A contribution associated with the self-energy is obtained from \eq{eegen} 
after picking out the summands with ${\bf p}' = {\bf k} +{\bf q}'$ and 
${\bf p}' = {\bf k} +{\bf q}$, i.e.
\beq
& & \langle [ \hat{\rho}^{\alpha}_{{\bf k} {\bf q}}, H_{{\rm int}}] 
\rangle^{\Sigma}  = \nonumber \\
& & = \frac{1}{2 {\cal V}} \sum_{{\bf q}',\beta} v_{{\bf q}'} \left\{ \langle 
c^{\dagger}_{{\bf k}+{\bf q}'} \sigma^{\beta} c_{{\bf k}+{\bf q}'}  \rangle 
\langle c^{\dagger}_{{\bf k}} \sigma^{\beta} \sigma^{\alpha} 
c_{{\bf k}+{\bf q}} \rangle \right. \nonumber \\
& & \left.  \qquad - \langle  c_{{\bf k}}^{\dagger} \sigma^{\alpha} 
\sigma^{\beta} c_{{\bf k}+{\bf q}} \rangle \langle c^{\dagger}_{{\bf k}+{\bf q}
 +{\bf q}'}  \sigma^{\beta} c_{{\bf k}+{\bf q}+{\bf q}'} \rangle   \right\}  
\label{self} \\
& & =  - \sum_{\beta}  \left\{ \Sigma^{\beta}_{{\bf k}}  \langle 
c^{\dagger}_{{\bf k}} \sigma^{\beta} \sigma^{\alpha} c_{{\bf k}+{\bf q}} 
\rangle  - \Sigma^{\beta}_{{\bf k}+{\bf q}}  \langle  c_{{\bf k}}^{\dagger} 
\sigma^{\alpha} \sigma^{\beta} c_{{\bf k}+{\bf q}} \rangle \right\} , 
\nonumber 
\eeq
where 
\beq
\Sigma^{\beta}_{{\bf k}} &=& - \frac{1}{2 {\cal V}} \sum_{{\bf q}'} 
v_{{\bf q}'-{\bf k}} \rho^{\beta}_{{\bf q}'0} \nonumber \\
&=& - \frac{1}{4 {\cal V}} \sum_{{\bf q}',\mu'} v_{{\bf q}'-{\bf k}} 
f_{{\bf q}'}^{\mu'} {\rm Tr} [(1+\mu' h^R_{{\bf q}'}) \sigma^{\beta}].
\eeq
The self-energy  $\Sigma^{\beta}_{{\bf k}}$ modifies the single-particle 
Hamiltonian $E_{{\bf k}}^{\beta} \to E_{{\bf k}}^{\beta} 
+\Sigma_{{\bf k}}^{\beta}$. After diagonalization we obtain the renormalized 
eigenvalues $\epsilon_{{\bf k}}^{\mu} \to \epsilon_{{\bf k}}^{\mu} + 
\Sigma_{{\bf k}}^{\mu}$,  
\be
\Sigma_{{\bf k}}^{\mu} =  - \frac{1}{2 {\cal V}} \sum_{{\bf q}',\mu'} 
v_{{\bf q}'-{\bf k}} f_{{\bf q}'}^{\mu'} 
[1+\mu \mu' \cos (\phi_{\bf k} -\phi_{{\bf q}'})], 
\ee
which correspond to the same eigenstates \eq{eigen}. Besides the shift of a  
chemical potential, the spectrum renormalization results in an effective value 
of the  Rashba splitting, which  has been previously studied in 
Ref.~\onlinecite{ghchen}. In our consideration we will, however, neglect this 
effect and discard the contribution \eq{self}.

The Hubbard's exchange term is given by the summands in \eq{eegen} with 
${\bf p}' = {\bf k}+{\bf q}+{\bf q}'$ and ${\bf p}' = {\bf k}$, i.e.
\beq
& & \langle [ \hat{\rho}^{\alpha}_{{\bf k} {\bf q}}, H_{{\rm int}}] 
\rangle^{{\rm Hub}} = \nonumber \\
& & = \frac{1}{2 {\cal V}} \sum_{{\bf q}',\beta} v_{{\bf q}'} \left\{  
\langle c^{\dagger}_{{\bf k}+{\bf q}'} \sigma^{\beta} c_{{\bf k}+{\bf q}
+{\bf q}'}  \rangle \langle c^{\dagger}_{{\bf k}+{\bf q}} \sigma^{\beta} 
\sigma^{\alpha} c_{{\bf k}+{\bf q}} \rangle \right. \nonumber \\
& & \qquad \quad \left. - \langle  c_{{\bf k}}^{\dagger} \sigma^{\alpha} 
\sigma^{\beta} c_{{\bf k}} \rangle \langle c^{\dagger}_{{\bf k}+{\bf q}'}  
\sigma^{\beta} c_{{\bf k}+{\bf q}+{\bf q}'} \rangle   \right\} \nonumber  \\
& & =  -\frac{1}{2 {\cal V}} \sum_{{\bf k}',\beta} v_{{\bf k}' -{\bf k}} 
{\cal B}_{{\bf k} {\bf q}}^{\alpha \beta} \rho^{\beta}_{{\bf k}' {\bf q}} 
\nonumber \\
& & \approx -\frac{\bar{v}_{{\bf q}}}{2 {\cal V}} \sum_{{\bf k}',\beta} 
{\cal B}_{{\bf k} {\bf q}}^{\alpha \beta} \rho^{\beta}_{{\bf k}' {\bf q}} 
= -\frac{\bar{v}_{{\bf q}}}{2 {\cal V}} \sum_{\beta} {\cal B}_{{\bf k} 
{\bf q}}^{\alpha \beta} \rho^{\beta}_{{\bf q}},
\label{hub}
\eeq
where $\bar{v}_{\bf q}$ is approximately regarded at small $q$ as a constant 
$\equiv 2 J = \frac{2 \pi e^2}{k_F}$. 
 
Collecting the contributions \eq{scr} and \eq{hub} and adding them to 
\eq{ebare}, we obtain the equation
\be
{\cal A}_{{\bf k} {\bf q}} \rho_{{\bf k} {\bf q}} = \frac{1}{{\cal V}} 
{\cal B}_{{\bf k} {\bf q}} \left(V_{{\bf q}} + F \rho_{{\bf q}} 
\right) 
\ee
where $F = {\rm diag} \{ v_{{\bf q}} -J, -J,-J,-J \}$. Solving it, we find 
that $\rho_{{\bf q}} = \tilde{\chi} V_{{\bf q}}$, where
\be
\tilde{\chi} = (1 -\chi F)^{-1} \chi 
\label{renchi}
\ee
is a density response matrix of the interacting system in the Hubbard's 
approximation.

\section{Explicit expressions for the functions $\chi^{(j)}$}
\label{toucal}

Let us list the functions $\check{f}_i^{(j)}$, $ C_i^{(j)}$ and 
$\delta_i^{(j)}$ which determine the response functions $\chi^{(j)}$ in  
Eqs. \eq{gunf} and \eq{funf}:
\beq
\check{f}_i^{(1)} &=& 0, \quad C_i^{(1)} = \frac{v -\mu y}{2 v^2 z}, 
\nonumber \\ 
\delta_i^{(1)} &=& \frac{(z^2 - \lambda w) -\mu y v}{(v -\mu y) z};  
\label{f1} \\
\check{f}_i^{(2)} &=& \frac{\lambda (\mu-y)}{2 z}, \,\, C_i^{(2)} 
= \frac{\mu \lambda (\lambda  w +\beta_i z v + \mu y v-z^2)}{2 v^2 z^2},
\nonumber \\
\delta_i^{(2)} &=& \frac{( \gamma_i v  +\mu y) z}{\lambda  w + \beta_i z v 
+ \mu y v-z^2}; \\
\check{f}_i^{(3)} &=& \frac{y (y-\mu)}{2 z^2}, \quad C_i^{(3)} 
= \frac{v z-\mu y z+\beta_i \mu y v}{2 v^2 z^2} , \nonumber \\
\delta_i^{(3)} &=& \frac{z^2 - \lambda w + \gamma_i \mu y v}{v z-\mu y z
+\beta_i \mu y v} ; \\
\check{f}_i^{(4)} &=& 0, \quad C_i^{(4)} = \frac{v +\mu y}{2 v^2 z}, 
\nonumber \\ 
\delta_i^{(4)} &=& \frac{z^2 - \lambda w}{(v +\mu y) z};  \\
\check{f}_i^{(5)} &=& \frac{\mu-y}{2 z}, \quad C_i^{(5)} 
= -\frac{ \mu (\lambda  w + \beta_i z v-z^2) }{2 v^2 z^2}, \nonumber \\
\delta_i^{(5)} &=& \frac{(\gamma_i  v -\mu y) z}{\lambda w + \beta_i z v -z^2}.
\label{f5}
\eeq
Employing \eq{f1}-\eq{f5} in the framework of the computational scheme 
elaborated in Ref.~\onlinecite{pletgr}, one can derive Eqs.~\eq{impi} and 
\eq{realpi}. 

The actual limits of integration in the latter expressions require further  
detailing, since the corresponding integrands ${\cal L}^{(j)} (\tau )$ \eq{lt} 
and  ${\cal R}^{(j)} (\tau )$ \eq{rt} are defined at $P (\tau ) >0$ and 
$P (\tau ) < 0$, respectively. The polynomial $P (\tau )$ \eq{pt} has the 
roots $\tau_k$ \eq{tauk}, which can be ordered differently depending on the 
values of $w$, $z$, and $y$.

\begin{figure}[t]
\includegraphics[width =5.25cm]{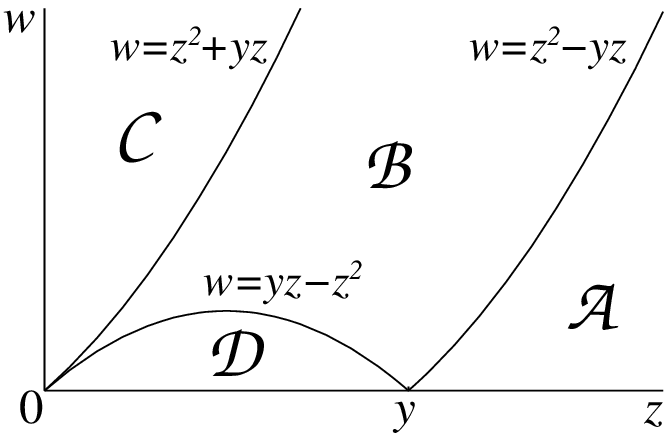}
\caption{The domains ${\cal A}$, ${\cal B}$, ${\cal C}$, and ${\cal D}$ 
corresponding to the different orderings \eq{orda}-\eq{ordd} of the roots 
$\tau_k$ \eq{tauk}.  }
\label{domain4}
\end{figure}

We identify the domains ${\cal A}$, ${\cal B}$, ${\cal C}$, and ${\cal D}$ 
in the plane $(z,w)$ [see Fig.~\ref{domain4}] such that
\beq
{\cal A} &:& \quad \tau_4 < \tau_2 < \tau_1 <\tau_3, \label{orda}\\
{\cal B} &:& \quad \tau_4 < \tau_2 < \tau_3 < \tau_1, \label{ordb}\\
{\cal C} &:& \quad  \tau_2 < \tau_4 < \tau_3 < \tau_1 , \label{ordc} \\
{\cal D} &:& \quad \tau_4 < \tau_3< \tau_2 < \tau_1 \label{ordd}.
\eeq
We also define the unit-step functions $\Theta ({\cal A}) = 
\Theta (z^2-y z -w)$, $\Theta ({\cal B}) = \Theta (w-z^2+y z) 
\Theta (z^2 + y z -w) \Theta( w+z^2 -y z)$, $\Theta ({\cal C}) 
= \Theta (w-z^2 -y z)$, and $\Theta ({\cal D}) =  (y z -z^2 -w)$, 
which realize projections onto these domains. Using these definitions 
we make the following decomposition 
\beq
\chi^{(j)} &=& \check{\chi}^{(j)}+ \Theta ({\cal A}) \chi_a^{(j)} 
+\Theta ({\cal B}) \chi_b^{(j)} \nonumber \\
&+& \Theta ({\cal C}) \chi_c^{(j)} + \Theta ({\cal D}) \chi_d^{(j)} ,
\eeq
where $\check{\chi}^{(j)}$ is nonzero only for $j=3$ and $j=5$:
\be
\check{\chi}^{(3)} = -2 \nu y^2/z^2, \quad \check{\chi}^{(5)} = 2 \nu y/z.
\label{cheq}  
\ee

Inside of each domain the root's ordering is fixed, and one has to find out 
how $\tau_{\sigma \lambda} (\ldots)$ occurring in Eqs. \eq{impi} and 
\eq{realpi} are arranged among $\tau_1, \ldots, \tau_4 $. 

Let us introduce $x_4< x_3 < x_2 < x_1$, where $x_k$'s are  identified 
with $\tau_k$'s differently in each domain according to \eq{orda}-\eq{ordd}, 
and define the primitives for the imaginary
\beq
A_1^{(j)+} (x) &=& \int_{x_1}^x dx' {\cal L}^{(j)} (x'), \quad x_1 <x, 
\label{pa1p} \\ 
A_2^{(j)} (x) &=& \int_{x_3}^x dx' {\cal L}^{(j)} (x'), 
\quad x_3 < x < x _2,\\  
A_1^{(j)-} (x) &=& -\int_x^{x_4} dx' {\cal L}^{(j)} (x'), \quad x<x_4 ,
\label{pa1m}
\eeq
and the real parts
\beq
B_1^{(j)} (x) &=& \int_{x_2}^x dx' {\cal R}^{(j)} (x'), \quad  x_2 <x<x_1, \\
B_2^{(j)} (x) &=& \int_{x_4}^x dx' {\cal R}^{(j)} (x'),  \quad x_4 <x<x_3.
\label{pb2}
\eeq
They have to be further detailed in each domain as well. For example, 
in the domain ${\cal C}$ we have $x_3 = \tau_4 \equiv  -z-y$, and therefore 
\be
 A_{2}^{(j)} (x) \to A_{2c}^{(j)} (x) = \int_{-z-y}^x dx' {\cal L}^{(j)} (x').
\label{a2c}
\ee
After such a specification the primitives \eq{pa1p}-\eq{pb2} can be explicitly 
found in terms of elliptic functions [see Appendix C of 
Ref.~\onlinecite{pletgr}]. In particular, we quote the $A_{2c}^{(j)}$ 
expressions for $j=2$ and $j=4$:
\beq
A_{2c}^{(2)} (x) = &-& \frac{k_c [(w/z-y)^2-z^2]}{4 z w \sqrt{w}} 
\left\{ 2 w F \left(\varphi_{2c} (x), k_c \right) \right. \nonumber \\
& & - (w/z)^2 n_{2c} \Pi (\varphi_{2c} (x), n_{2c}, k_c) \nonumber \\
& & - \left. (z^2-y^2) \widetilde{n}_{2c} \Pi (\varphi_{2c} (x), 
\widetilde{n}_{2c}, k_c) \right\} ,
\label{a0ye}
\eeq
\beq
A_{2c}^{(4)} (x) &=& \frac{w}{2 (z^2 -y^2) x} \sqrt{\frac{z x-w}{zx+w} 
[(x+y)^2-z^2]} \nonumber \\
&-& \frac{k_c ((w/z-y)^2 -z^2)}{4 \sqrt{w} (z^2 -y^2)}  
\left\{z F (\varphi_{2c} (x), k_c)   \right.\nonumber \\
& & - \left. y  \widetilde{n}_{2c}  \Pi (\varphi_{2c} (x), 
\widetilde{n}_{2c}, k_c) \right\} \nonumber \\
&+&  \frac{z \sqrt{w}}{k_c (z^2-y^2)}  E (\varphi_{2c} (x), k_c) ,
\label{azze}
\eeq
where
\beq
\varphi_{2c} (x) &=& \arcsin \sqrt{\frac{(x + y+z) (w/z-y+z)}{2 (z x +w)}}, \\
k_c &=& \frac{2 \sqrt{w}}{\sqrt{(z+w/z)^2 -y^2}} , \\
n_{2c} &=& \frac{2 z}{w/z+z-y}, \quad \widetilde{n}_{2c} = \frac{w}{z (z+y)} 
n_{2c}.
\eeq

Let us now establish the actual limits of integration in Eq.~\eq{impi}. 
Rewriting it in terms of \eq{pa1p}-\eq{pa1m}, we obtain ${\rm Im} \chi^{(j)}$ 
in every domain
\begin{widetext}
\beq
&-& \frac{1}{\nu} {\rm Im} \chi_{a}^{(j)} = -\Theta (w/z-t_1) 
\left[ A_{2a}^{(j)} (w/z) - A_{2a}^{(j)} (t_1) + A_{2a}^{(j)} (-t_1) \right] 
- \Theta (t_2-z-y)  A_{1a}^{(j)-} (-t_2 ) \nonumber \\
& & +\Theta (t_2-z+y) A_{1a}^{(j)+} (t_2) - \Theta (1 -4 w) 
\left\{  \Theta (t_4 - z+y) A_{1a}^{(j)+} (t_4) - 
\Theta (t_3-z+y) A_{1a}^{(j)+} (t_3) \right.  \nonumber \\
& & - \left. \Theta (w/z-t_3) \left[  A_{2a}^{(j)} (w/z) - A_{2a}^{(j)} (t_3) 
+ A_{2a}^{(j)} (-t_3) \right] - \Theta (t_4-z-y) A_{1a}^{(j)-} (-t_4) \right\},
 \label{icha}
\eeq

\beq
&-& \frac{1}{\nu} {\rm Im} \chi_{b}^{(j)} = -\Theta (z-y-t_1)  
\left[A_{2b}^{(j)} (z-y) -  A_{2b}^{(j)} (t_1) \right] 
- \Theta (z-y+t_1) A_{2b}^{(j)} (-t_1)  \nonumber  \\
& & - \Theta (y- z-t_1) A_{2b}^{(j)} (z-y) - \Theta (t_2 -z-y)   
A_{1b}^{(j)-} (-T_2) +  A_{1b}^{(j)+} (t_2) - \Theta (1-4 w) \times 
\nonumber \\
& & \times \left\{ -\Theta (t_4 -z-y) A_{1b}^{(j)-} (-t_4) 
+\Theta (t_4 - w/z) A_{1b}^{(j)+} (t_4) 
- \Theta (t_3 - w/z) A_{1b}^{(j)+} (t_3) \right. \nonumber \\
& & - \Theta (z-y-T_3) \left[ A_{2b}^{(j)} (z-y) - A_{2b}^{(j)} (t_3) \right]
 + \Theta (w/z-t_3) \Theta (z-y+t_3) \left[ A_{2b}^{(j)} (z-y) 
- A_{2b}^{(j)} (-t_3) \right] \nonumber \\ 
& & \left. - \Theta (w/z-t_4) \left[ A_{2b}^{(j)} (z-y) 
- A_{2b}^{(j)} (-t_4) \right] - \Theta (t_4 -w/z) \Theta (w/z-t_3) 
A_{2b}^{(j)} (z-y) \right\}, \label{ichb}
\eeq

\beq
&-& \frac{1}{\nu} {\rm Im} \chi_{c}^{(j)} =  \Theta (t_2 -w/z) 
\left[ A_{1c}^{(j)+} (t_2)  -  A_{1c}^{(j)-} (-t_2) \right] - 
\Theta (z-y-t_1) \left[ A_{2c}^{(j)} (z-y) - A_{2c}^{(j)} (t_1)  \right] 
\nonumber \\
& & - \Theta (z-y+t_1) \Theta ( z+y-t_1) A_{2c}^{(j)} (-t_1)
- \Theta (y-z-t_1) A_{2c}^{(j)} (z-y) - \Theta (1 - 4 w) \times \nonumber \\
& & \times  \left\{ \Theta (t_4 -w/z) \left[  A_{1c}^{(j)+} (t_4) 
-  A_{1c}^{(j)-} (-t_4) \right]   -\Theta (z-y-t_3) \left[ A_{2c}^{(j)} (z-y) 
-  A_{2c}^{(j)} (t_3)\right] \right. \nonumber \\
& & + \Theta (z+y-t_3) \Theta (z-y+t_3 ) \left[ A_{2c}^{(j)} (z-y) 
- A_{2c}^{(j)} (-t_3)\right] - \Theta (z+y-t_4) \Theta (z-y+t_4 ) \times 
\nonumber \\
& & \times \left. \left[ A_{2c}^{(j)} (z-y) -  A_{2c}^{(j)} (-t_4)\right] 
-  \Theta (t_4-z-y) \Theta (z+y-t_3) A_{2c}^{(j)} (z-y) \right\}, 
\label{ichc}
\eeq

\beq
- \frac{1}{\nu} {\rm Im} \chi_{d}^{(j)} =  A_{1d}^{(j)+} (t_2) 
- A_{1d}^{(j)-} (-t_2) + A_{1d}^{(j)-} (-t_4) - A_{1d}^{(j)+} (t_4) . 
\label{ichd}
\eeq
\end{widetext}
In the above expressions we have used the definitions
\beq
t_{1,2} &=& \frac12 \left[\sqrt{1 + 4 w} \mp 1 \right], \label{t12} \\  
t_{3,4} &=&  \frac{1}{2} \left[1 \mp \sqrt{1 - 4 w}\right] \label{t34}.
\eeq

The actual limits of integration for the real part \eq{realpi} can be found 
in a similar way. However, Eq.~\eq{realpi} accounts only the term ${\rm Re} 
\chi^{(j),I}$. Complementing it by  ${\rm Re} \chi^{(j),II}$ [see the 
remark 5) in the end of Sec.~\ref{operator}], we present the sum of the both 
terms in every domain 
\begin{widetext}
\beq
&-& \frac{1}{\nu} {\rm Re} \chi_{a}^{(j)} = \Theta (t_2 - z+y) B_{1a}^{(j)} 
(z-y) + \Theta (z-y-t_2) B_{1a}^{(j)} (t_2) 
+\Theta (t_1-w/z) B_{1a}^{(j)} (t_1) -2 B_{1a}^{(j)} (\tilde{t}_1) 
\nonumber \\
& & -  [2+\Theta (w/z-t_1)] B_{2a}^{(j)} (-w/z) - \Theta (t_1-w/z) 
B_{2a}^{(j)} (-T_1) + 2 B_{2a}^{(j)} (-\tilde{t}_2) - \Theta (z+y-t_2) 
B_{2a}^{(j)} (-t_2)  
\nonumber \\ 
& & + 2 \Theta (y^2 - 4 w) \left[ B_{2a}^{(j)} (\tilde{t}_3) 
+ B_{2a}^{(j)} (\tilde{t}_4) \right] +  2 \Theta (4 w - y^2) 
\left[B_{2a}^{(j)} (\tilde{t}_3^c) + B_{2a}^{(j)} (\tilde{t}_4^c) \right]  
+ \Theta (1 - 4 w) \left\{\Theta (z-y-t_4) B_{1a}^{(j)} (t_4) \right.  
\nonumber \\
& & + \left. \Theta (t_3 - w/z) \Theta (z-y-t_3 ) B_{1a}^{(j)} (t_3)
 + [\Theta (t_3 -z+y) +  \Theta (t_4 -z+y) ] B_{1a}^{(j)} (z-y) 
- \Theta (z+y-t_4 ) B_{2a}^{(j)} (-t_4) \right. \nonumber \\
& & - \left.  \Theta (t_3 -w/z) B_{2a}^{(j)} (-t_3) - \Theta (w/z-t_3) 
B_{2a}^{(j)} (-w/z)  \right\}  + \Theta (4 w -1) \left\{ B_{1a}^{(j)} (t_3^c) 
+  B_{1a}^{(j)} (t_4^c) - B_{2a}^{(j)} (-t_3^c) \right. \nonumber \\
& & - \left. B_{2a}^{(j)} (-t_4^c) \right\} + s^{(j)}_1 
\frac{\pi \sqrt{z^2-y^2}}{z} +\tilde{s}^{(j)}_1 \frac{\pi w}{z \sqrt{z^2-y^2}} 
-2 s^{(j)}_3 \left[ B_{1a}^{(j)} (z-y) - B_{2a}^{(j)} (-w/z) \right], 
\label{rcha}
\eeq

\beq
&-& \frac{1}{\nu} {\rm Re} \chi_{b}^{(j)} = - B_{1b}^{(j)} (w/z) +  
2 B_{1b}^{(j)} (\tilde{t}_1) - \Theta (t_1 -z+y) B_{1b}^{(j)} (t_1)
- \Theta (y -z-t_1) B_{1b}^{(j)} (-t_1)
- \Theta (z+y-t_2 ) B_{2b}^{(j)} (-t_2) \nonumber \\ 
& & + 2 B_{2b}^{(j)} (-\tilde{t}_2) - \Theta (w/z - t_1) B_{2b}^{(j)} (-w/z) 
- \Theta (t_1-w/z ) B_{2b}^{(j)} (-t_1) - 2  \Theta (y^2 - 4 w) \Theta (w-z^2)
\left[ B_{1b}^{(j)} (\tilde{t}_3) + B_{1b}^{(j)} (\tilde{t}_4 ) \right] 
\nonumber \\ 
& & + 2 \Theta (y^2 - 4 w) \Theta (z^2-w ) \left[ B_{2b}^{(j)} (\tilde{t}_3)
+ B_{2b}^{(j)} (\tilde{t}_4) - 2 B_{2b}^{(j)} (-w/z) \right] 
- 2 \Theta (4 w -y^2) \left[ B_{1b}^{(j)} (\tilde{t}_3^c) + B_{1b}^{(j)} 
(\tilde{t}_4^c)\right]
\nonumber \\
&  & +\Theta (1 - 4 w)\left\{ [\Theta (t_3-w/z) + \Theta (t_4 -w/z)] 
B_{1b}^{(j)} (w/z)+ \Theta (w/z - t_3) \Theta (t_3 -z+y) B_{1b}^{(j)} (t_3) 
+ \Theta (w/z- t_4) B_{1b}^{(j)} (t_4 ) \right. \nonumber \\
& & + [\Theta (t_3-w/z)+\Theta (t_4-w/z)] B_{2b}^{(j)} (-w/z) 
- \Theta (z+y-t_4 ) \Theta (t_4-w/z ) B_{2b}^{(j)} (-t_4)  + \Theta (y-z-t_3) 
B_{1b}^{(j)} (-t_3) \nonumber \\
& & - \left. \Theta (t_3-w/z) B_{2b}^{(j)} (-t_3) \right\} 
+ \Theta (4 w-1) \left\{ B_{1b}^{(j)} (t_3^c) +B_{1b}^{(j)} (t_4^c) + 
B_{1b}^{(j)} (-t_3^c) + B_{1b}^{(j)} (-t_4^c) \right\}+ \Theta (1 - 4 w) 
\Theta (y - z) \times \nonumber \\
& & \times  \left(s^{(j)}_1 \frac{\pi w}{z^2} + \tilde{s}^{(j)}_1 \pi - 
2 s^{(j)}_2 \left[B_{1b}^{(j)} (w/z) - B_{2b}^{(j)} (-w/z) \right] \right)  
- \left[\Theta (1 - 4 w) \Theta (4 w - y^2) -  \Theta (y^2 - 4 w)\right] 
\Theta (z - y) \times \nonumber \\
& & \times  \left( s_1^{(j)} \frac{\pi \sqrt{z^2 - y^2}}{z} + 
\tilde{s}^{(j)}_1 \frac{\pi w}{z \sqrt{z^2 - y^2}} - 2 s_3^{(j)} 
\left[ B_{1b}^{(j)} (w/z) - B_{2b}^{(j)} (-w/z) \right] \right), 
\label{rchb}
\eeq

\beq
&-& \frac{1}{\nu} {\rm Re} \chi_{c}^{(j)} =2 B_{1c}^{(j)} (\tilde{t}_1) 
-  \Theta (t_2 - w/z) B_{1c}^{(j)} (w/z) -\Theta (w/z-t_2) B_{1c}^{(j)} (t_2) 
- \Theta (t_1 -z+y) B_{1c}^{(j)} (t_1)  \nonumber \\
& & - \Theta (y -z-t_1) B_{1c}^{(j)} (-t_1)  + \Theta (z+y-t_1) B_{2c}^{(j)} 
(-z-y) + \Theta (t_1 -z-y) B_{2c}^{(j)} (-t_1) +\Theta (w/z-t_2) B_{2c}^{(j)} 
(-t_2) \nonumber \\
& & - 2 B_{2c}^{(j)} (-\tilde{t}_2) -2 \Theta (y^2 - 4 w) \left[ B_{1c}^{(j)} 
(\tilde{t}_3) + B_{1c}^{(j)}  (\tilde{t}_4) \right] - 2 \Theta (4 w -y^2) 
\left[ B_{1c}^{(j)} (\tilde{t}_3^c) + B_{1c}^{(j)} (\tilde{t}_4^c)\right] + 
\Theta (1 - 4 w) \times \nonumber \\
& & \times \left\{ \Theta (t_4 - w/z) B_{1c}^{(j)} (w/z) + \Theta (w/z-t_4 ) 
B_{1c}^{(j)} (t_4) +  \Theta (t_3 - z+y) B_{1c}^{(j)} (t_3)+ \Theta (y- z-t_3) 
B_{1c}^{(j)} (-t_3) \right. \nonumber \\
& & + [\Theta (t_4-z-y) + \Theta (t_3-z-y)] B_{2c}^{(j)} (-z-y) - 
\Theta (w/z- t_4 ) \Theta (t_4-z-y) B_{2c}^{(j)} (-t_4) \nonumber \\
& & -\left.  \Theta (t_3-z-y) B_{2c}^{(j)} (-t_3)  \right\} +  \Theta (4 w -1) 
\left\{ B_{1c}^{(j)} (t_3^c) + B_{1c}^{(j)} (t_4^c)+ B_{1c}^{(j)} (-t_3^c)+ 
B_{1c}^{(j)} (-t_4^c)\right\} \nonumber \\
& & +\Theta (1 - 4 w) \Theta (y - z) \left(s^{(j)}_1 \frac{\pi w}{z^2} + 
\tilde{s}^{(j)}_1 \pi - 2 s^{(j)}_2 \left[B_{1c}^{(j)} (w/z) - B_{2c}^{(j)} 
(-z-y) \right] \right)- \Theta (1 - 4 w) \Theta (z - y) \times \nonumber \\
& & \times  \left(s^{(j)}_1 \frac{\pi \sqrt{z^2 - y^2}}{z} + 
\tilde{s}^{(j)}_1 \frac{\pi w}{z \sqrt{z^2 - y^2}} - 2 s_3^{(j)} 
\left[ B_{1c}^{(j)} (w/z) - B_{2c}^{(j)} (-z-y) \right] \right),
\label{rchc} 
\eeq

\beq
&-& \frac{1}{\nu} {\rm Re} \chi_{d}^{(j)} = 2 B_{1d}^{(j)} (\tilde{t}_1)
- B_{1d}^{(j)} (t_1)-B_{1d}^{(j)} (-t_1) -2 B_{2d}^{(j)} (z-y)+ 2 B_{2d}^{(j)} 
(-\tilde{t}_2) -2 B_{1d}^{(j)} (\tilde{t}_4)\nonumber \\
& & + B_{1d}^{(j)} (t_3)+B_{1d}^{(j)} (-t_3) +2 B_{2d}^{(j)} (\tilde{t}_3) 
+ s^{(j)}_1 \frac{\pi w}{z^2} +\tilde{s}^{(j)}_1 \pi - 2 s_2^{(j)} 
\left[B_{1d}^{(j)} (w/z) - B_{2d}^{(j)} (z-y)\right].  
\label{rchd}
\eeq
\end{widetext}

In these expressions we have used along with \eq{t12} and \eq{t34} 
the following arguments
\beq
\tilde{t}_{1,2} &=& \frac12 \left[ \sqrt{y^2 + 4 w} \mp y \right],  \\
\tilde{t}_{3,4} &=& \frac12 \left[ - y \mp \sqrt{y^2 - 4 w}\right] , \\
t_{3,4}^c &=& \frac12[1 \mp i \sqrt{4 w -1}], \\
\tilde{t}_{3,4}^c &=& \frac12 [- y \mp i \sqrt{4 w -y^2} ].
\eeq
Note that $t_{3,4}^c$ and $\tilde{t}_{3,4}^c$ are complex-valued, which assumes
the analytic continuation of  elliptic functions hinted in Appendix B of 
Ref.~\onlinecite{pletgr}. 

Transforming ${\rm Re} \chi^{(j),II}$ into the representation in terms of 
$B_1^{(j)}$ and $B_2^{(j)}$, we accumulate the residue terms which are 
different for each response function. They are fully defined by the following 
sets of constants
\beq
s_1^{(1)} &=& \tilde{s}^{(1)}_1 =s^{(1)}_2=s^{(1)}_3 =0, \\
s^{(2)}_1 &=& s^{(2)}_2 =1, \quad \tilde{s}^{(2)}_1=s^{(2)}_3 =0, \\
s^{(3,4)}_2 &=& s^{(3,4)}_3 =1, \quad s^{(3,4)}_1=\tilde{s}^{(3,4)}_1 =0, \\
\tilde{s}^{(5)}_1 &=& s_2^{(5)} =1, \quad s_1^{(5)} = s_3^{(5)} =0.
\eeq

We recall once again that for $j=3$ and $j=5$ Eqs.~\eq{rcha}-\eq{rchd} must be 
complemented by $\check{\chi}^{(j)}$ \eq{cheq}.

\end{document}